%% file: main_arxiv.tex
\documentclass[runningheads]{llncs}

\usepackage[T1]{fontenc}
\usepackage{graphicx}
\usepackage{amsmath}
\usepackage{booktabs}
\usepackage{subcaption}
\usepackage{hyperref}
\usepackage{xcolor}
\usepackage{url}
\usepackage[most]{tcolorbox}
\usepackage{enumitem}
\usepackage{float}
\usepackage{tabularx}
\usepackage{colortbl}
\usepackage{makecell}
\usepackage{siunitx}
\usepackage{multirow}
\usepackage{marvosym}

\definecolor{boxrule}{RGB}{180,180,180}
\definecolor{tabgray}{RGB}{245,245,245}
\definecolor{boxgray}{RGB}{230,230,230}

\makeatletter
\renewcommand{\paragraph}{%
  \@startsection{paragraph}{4}%
  {\z@}{0ex \@plus 1ex \@minus .2ex}{-1em}%
  {\normalfont\normalsize\bfseries}%
}
\makeatother
\newcommand{\Paragraph}[1]{\paragraph{{#1.}}}

\title{DrawSim-PD: Simulating Student Science Drawings to Support NGSS-Aligned Teacher Diagnostic Reasoning}
\titlerunning{DrawSim-PD}

\author{
Arijit Chakma\inst{1} \and
Peng He\inst{3} \and
Honglu Liu\inst{2,3} \and 
Zeyuan Wang\inst{3} \and 
Tingting Li\inst{3} \and \\\vspace{1mm}
Tiffany D. Do\inst{1} \and
Feng Liu\inst{1}\textsuperscript{\Letter}
}

\authorrunning{A. Chakma et al.} 

\institute{
Department of Computer Science, Drexel University \\
\and
College of Chemistry, Beijing Normal University
\and
Department of Teaching and Learning, Washington State University \\
}

\begin{document}
\maketitle

\begingroup
\renewcommand{\thefootnote}{} 
\footnotetext{\Letter\ Corresponding author. \texttt{fl397@drexel.edu}}
\endgroup

\input{sec_arXiv/abstract_0201}

\input{sec_arXiv/intro_0201}
\input{sec_arXiv/related_work_0201}
\input{sec_arXiv/methods_0201}
\input{sec_arXiv/exp_0201}
\input{sec_arXiv/conclusion_0201}
\bibliographystyle{splncs04}
\bibliography{references}

\appendix
\input{sec_arXiv/appendix_0201}

\end{document}

%% file: sec_arXiv/abstract_0201.tex
\begin{abstract}
Developing expertise in diagnostic reasoning requires practice with diverse student artifacts, yet privacy regulations prohibit sharing authentic student work for teacher professional development (PD) at scale.
We present \textbf{DrawSim-PD}, the first generative framework that 
simulates NGSS-aligned, student-like science drawings exhibiting \emph{controllable 
pedagogical imperfections} to support teacher training. Central to our approach 
are \emph{capability profiles}---structured cognitive states encoding what 
students at each performance level can and cannot yet demonstrate. These profiles 
ensure cross-modal coherence across generated outputs: (i) a student-like 
drawing, (ii) a first-person reasoning narrative, and (iii) a teacher-facing 
diagnostic concept map. Using 100 curated NGSS topics spanning K--12, we 
construct a corpus of 10,000 systematically structured artifacts. Through an 
expert-based feasibility evaluation, K--12 science educators verified the 
artifacts' alignment with NGSS expectations ($>$84\% positive on core items) 
and utility for interpreting student thinking, while identifying refinement 
opportunities for grade-band extremes. We release this open infrastructure to 
overcome data scarcity barriers in visual assessment research.
\href{https://vilab-group.com/project/drawsim-pd}{Project}

\keywords{K--12 science education \and formative assessment \and teacher diagnostic reasoning \and student drawings \and NGSS \and generative AI}
\end{abstract}

%% file: sec_arXiv/intro_0201.tex
\begin{figure}[t]
    \centering
    \includegraphics[width=1\textwidth]{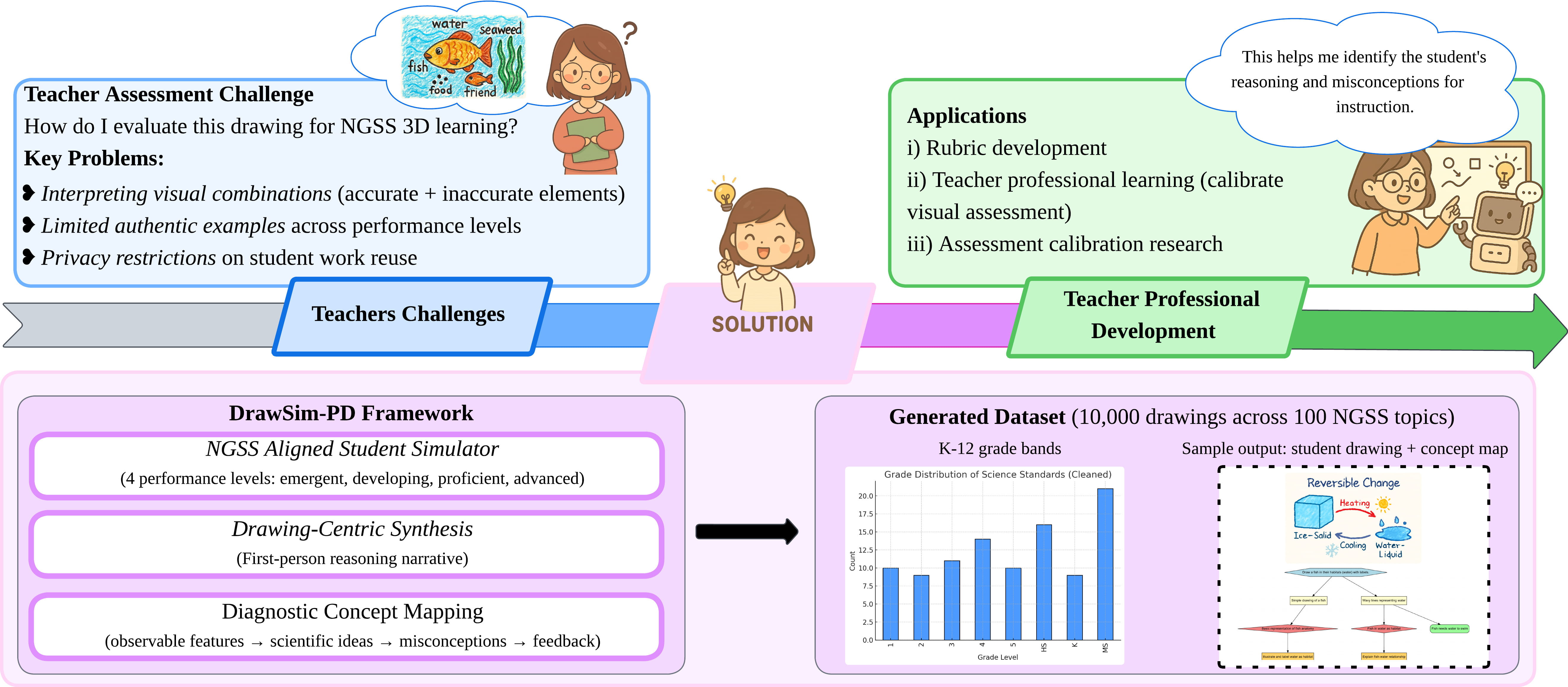}
\caption{\textbf{The DrawSim-PD Framework.} (Left) Teachers struggle to practice diagnostic reasoning due to a scarcity of privacy-compliant student drawings. (Center) The system uses \emph{capability profiles} to bridge the gap, generating student-like artifacts with controlled misconceptions. (Right) The output serves as scalable infrastructure for teacher calibration and professional development.}
    \vspace{-4mm}
    \label{fig:teaser}
\end{figure}

\section{Introduction}

\textbf{\textit{How can teachers develop expertise in diagnosing student understanding from hand-drawn scientific representations when the drawings they need for practice are either unavailable or unusable?}}

A fifth-grade class sketches the water cycle. Some drawings show only 
evaporation; others include arrows looping in the wrong direction. Under the 
Next Generation Science Standards (NGSS)~\cite{NGSS}, teachers must interpret 
such drawings as evidence of three-dimensional (3D) learning, integrating 
Disciplinary Core Ideas (DCIs), Science and Engineering Practices (SEPs), and 
Crosscutting Concepts (CCCs). Diagnosing student 3D understanding is a core component of 
pedagogical content knowledge~\cite{Shulman1986} and central to effective 
formative assessment~\cite{BlackWiliam1998}. Yet, teachers rarely have access 
to diverse, grade-appropriate drawing exemplars to calibrate their judgments, 
both because authentic student work is hard to collect at scale and because 
privacy protections limit reuse~\cite{Heritage2010}.

Research on teacher noticing emphasizes that expertise develops through 
repeated exposure to student thinking~\cite{Sherin2011,vanEs2011}. Professional 
development programs use student work samples to help teachers recognize 
patterns of understanding and misconception~\cite{Kazemi2009}. However, for 
visual science assessment, such collections are sparse. Recent advances in 
generative AI offer a potential solution: synthesizing student-like artifacts 
without relying on actual student data. For example, systems like Generative 
Students~\cite{GenerativeStudents2024} simulate student profiles for text-based 
assessment, while TutorUp~\cite{pan2025tutorup} and TeachTune~\cite{Jin2025} 
generate dialogue interactions for teacher training. However, these focus on 
text; diagram generators like DiagrammerGPT~\cite{Zala2023DiagrammerGPT} 
produce accurate illustrations but fail to simulate developmental constraints 
and authentic errors. To our knowledge, no framework generates educationally 
plausible student drawings reflecting both scientific ideas and realistic 
limitations at different performance levels.

This reflects a significant technical challenge. The goal is not merely generating diagrams—systems like DiagrammerGPT excel at this. The challenge lies in generating \emph{pedagogically meaningful imperfection}: partial understandings, spatial reasoning errors, and conceptual gaps that render student work diagnostically informative. This requires inverting the generative AI objective, systematically introducing educationally plausible errors while maintaining coherence across visual, textual, and conceptual modalities. We address this gap with \textbf{DrawSim-PD} (Fig.~\ref{fig:teaser}), a framework that simulates student-like science drawings to support teacher professional development (\textbf{PD}) and diagnostic reasoning. Central to our approach are \emph{capability profiles} derived from 
NGSS performance expectations encoding what students at each level can and 
cannot yet demonstrate. These profiles guide joint generation of: (i)~a 
student-like drawing, (ii)~a first-person reasoning narrative, and (iii)~a 
teacher-facing \emph{diagnostic concept map} linking observable features to 
underlying understanding and misconceptions.

DrawSim-PD comprises three modules: (1)~\emph{NGSS-Aligned Student Simulator} 
decomposes performance expectations into capability profiles; (2)~\emph{Drawing-Centric Synthesis} generates narratives and drawings conditioned on 
profiles; (3)~\emph{Diagnostic Concept Mapping} produces four-layer concept 
maps linking observations to understanding and instructional next steps. We 
constructed a corpus across 100 NGSS topics with systematic variation across 
four performance levels and K--12 grade bands.

\textbf{Broader Impact.} DrawSim-PD enables three previously impractical applications: (i)~\emph{scalable calibration exercises} where teachers evaluate synthetic drawings and compare judgments without compromising student privacy; (ii)~\emph{targeted misconception libraries} for teacher educators to curate examples of specific conceptual difficulties across performance levels; (iii)~\emph{assessment research infrastructure} providing unlimited samples for studying diagnostic reasoning, circumventing privacy constraints that limit visual assessment research. Our contributions:

$\diamond$ We introduce a capability-profile mechanism that enables the generation of student-like drawings with systematically varied misconceptions, achieving controllable pedagogical imperfection aligned to curriculum standards.

$\diamond$ We devise an automated diagnostic scaffolding module that transforms visual artifacts into structured teacher supports via generated diagnostic concept maps.

$\diamond$ We release a 10,000-artifact corpus with structured metadata as open research infrastructure, representing the largest collection of curriculum-aligned student drawing simulations to date.

$\diamond$ We validate the system's pedagogical fidelity through an expert feasibility study, where experienced educators confirmed that the generated outputs are NGSS-aligned and pedagogically authentic.

%% file: sec_arXiv/related_work_0201.tex
\section{Related Work}

\subsection{Teacher Diagnostic Reasoning and Professional Development}

Effective science teaching requires diagnostic reasoning: the capacity to interpret student work, identify underlying understanding and misconceptions, and respond instructionally~\cite{Shulman1986,BlackWiliam1998}. Research on teacher noticing highlights that this expertise develops through sustained engagement with diverse examples of student thinking~\cite{Sherin2011,vanEs2011}. However, this ``professional vision'' is not innate; it must be cultivated. Professional development programs that center on collaborative analysis of student work have proven effective for building interpretive skill~\cite{Kazemi2009}, allowing teachers to move from merely describing student work to interpreting the cognitive processes behind it.
For these programs to be effective, they require ``calibration sets''—collections of artifacts that represent a range of proficiency levels and specific conceptual pitfalls. Such activities rely on well-structured exemplars, yet access to diverse, systematically organized collections remains limited. This scarcity is particularly acute for visual artifacts, where strict privacy concerns regarding student handwriting and drawing styles restrict sharing and reuse at scale~\cite{LiuSynData2024}.
Student drawings pose particular challenges for diagnostic reasoning. Unlike written responses, drawings possess a high degree of representational ambiguity. They embed scientific understanding within spatial arrangements, symbolic conventions (\emph{e.g.}, arrows, labels), and developmental drawing skills that vary significantly by age and experience~\cite{Ainsworth2011,Quillin2015}. Teachers often struggle to distinguish between a student's lack of scientific knowledge and a lack of artistic ability or motor control. Consequently, teachers must simultaneously evaluate scientific accuracy, identify misconceptions, and calibrate expectations to grade-appropriate norms. This interpretive complexity, combined with the scarcity of usable training data, motivates our focus on generating diverse, NGSS-aligned drawing exemplars with explicit diagnostic scaffolding.

\subsection{NGSS-Aligned Assessment and Student Simulation}

The Next Generation Science Standards (NGSS) require integrated 3D learning across Disciplinary Core Ideas (DCIs), Science and Engineering Practices (SEPs), and Crosscutting Concepts (CCCs)~\cite{NGSS}. This integration creates significant assessment challenges: teachers must interpret complex artifacts that combine scientific content, reasoning practices (modeling), and conceptual connections~\cite{Laverty2016,Fulmer2018}. Prior research indicates that teachers report difficulty consistently identifying reasoning patterns within student drawings, especially when representations combine accurate elements with developmental limitations or partial understanding.
Recent work explores large language models (LLMs) for simulating student behaviors to support teacher training. Lu and Wang~\cite{GenerativeStudents2024} introduced \textit{Generative Students}, using LLM-simulated student profiles to evaluate assessment items, demonstrating that synthetic learner responses can approximate authentic student variation. This line of work has rapidly expanded: TutorUp~\cite{pan2025tutorup} generates realistic dialogue interactions for novice tutor training, while TeachTune~\cite{Jin2025} enables teachers to test pedagogical agents against diverse simulated profiles. Beyond one-on-one interactions, systems have been developed to simulate student behaviors in classroom discussions~\cite{zhang2024simulating}, populate virtual classrooms with diverse agents~\cite{ZHENG2025100990}, and model specific metacognitive processes~\cite{li2025exploring}. Furthermore, Wu et al.~\cite{wu2025embracing} have explored methods for simulating diverse cognitive levels to capture realistic student imperfections.
However, existing student simulation remains predominantly text-based, focusing on dialogue and written responses rather than visual reasoning. This is a critical limitation for science education, where modeling—often expressed visually—is a primary practice. While automated NGSS scoring systems classify student reasoning in text~\cite{liu_ai_diagnosis_2024,kaldaras2022rubric} and commercial tools like Flint generate aligned textual assessments~\cite{flint_ngss_assessments}, existing work has not yet provided teachers with diverse, authentic \textit{visual} exemplars for developing diagnostic expertise in visual science assessment.

\subsection{Synthetic Educational Data and Diagram Generation}

Synthetic data generation addresses privacy constraints and enables scalable educational research under regulations like FERPA. Student drawings present heightened privacy concerns, as handwriting styles and visual conventions can be personally identifiable even when anonymized. Educational applications have employed Bayesian networks~\cite{Singh2023} and Markov chains~\cite{Vie2022}, though these often capture narrow modalities and underrepresent classroom diversity~\cite{LiuSynData2024,LiuSynData2025}. More recently, multimodal frameworks like MAGID~\cite{aboutalebi-etal-2024-magid} and SMMQG~\cite{wu-etal-2024-synthetic} have demonstrated potential for generating multimodal datasets, though they frequently lack the strict curriculum alignment and authentic student reasoning patterns required for teacher training.
In the domain of visual generation, diagram generation has advanced with systems like DiagrammerGPT~\cite{Zala2023DiagrammerGPT}, which creates scientifically accurate diagrams through LLM planning and layout optimization. However, a fundamental tension exists between standard generative goals and educational simulation. Systems like DiagrammerGPT or DALL-E 3 prioritize technical accuracy and aesthetic quality. They are optimized to follow a prompt perfectly (\emph{e.g.}, ``Draw a correct water cycle''). In contrast, simulating student work requires ``controllable imperfection'': producing sketches that are developmentally appropriate (\emph{e.g.}, using crayon textures, uneven lines) and contain specific, plausible misconceptions (\emph{e.g.}, missing gravity, broken cycles).
This gap between technical accuracy and educational authenticity motivates our work. Teachers need synthetic student drawings that reflect realistic performance variation for calibration and professional development. Existing approaches generate modalities independently, failing to maintain coherence across visual and textual representations (\emph{e.g.}, a text claiming a student understands ``precipitation'' while the image omits it). DrawSim-PD addresses this by jointly generating drawings, reasoning narratives, and diagnostic concept maps conditioned on shared capability profiles derived from NGSS performance expectations.

%% file: sec_arXiv/methods_0201.tex
\begin{figure*}[t]
    \centering
    \includegraphics[width=\textwidth]{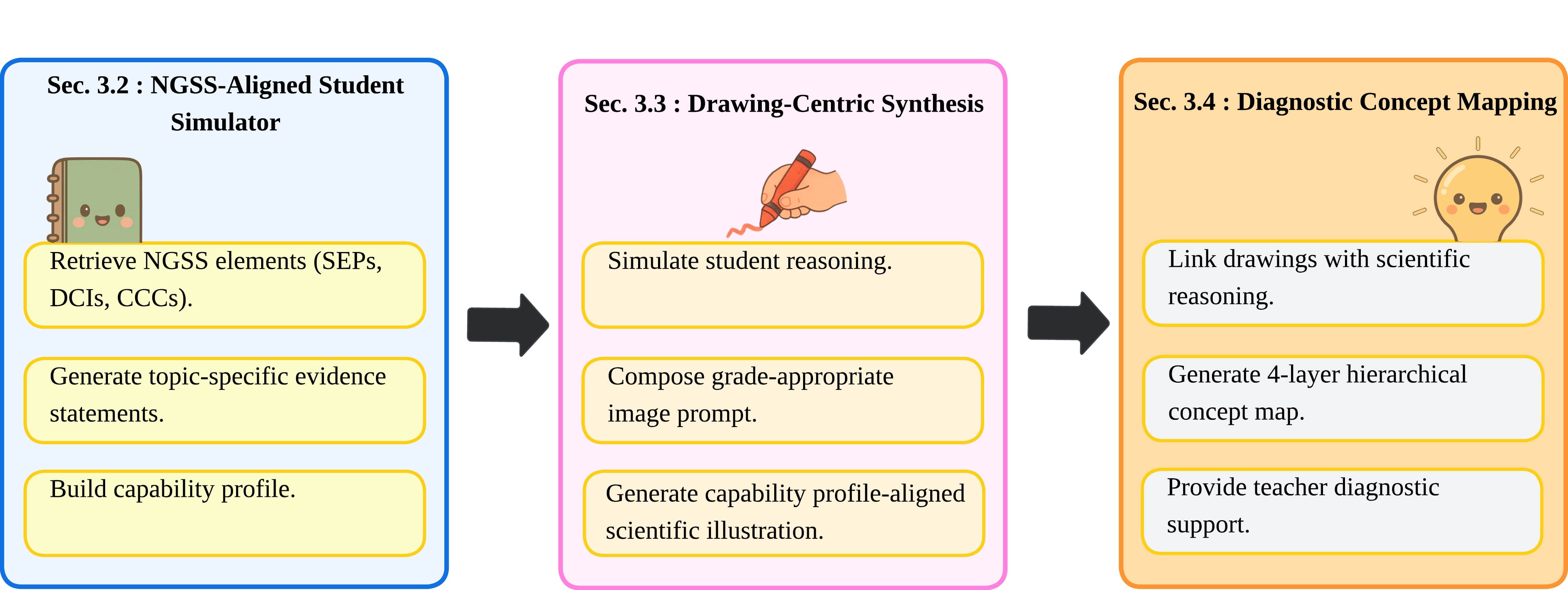} 
    \caption{The \textbf{DrawSim-PD} framework comprises three modules: 
    (1) \textbf{NGSS-Aligned Student Simulator}, which generates topic-specific 
    evidence statements and capability profiles representing diverse K--12 student 
    performance levels; 
    (2) \textbf{Drawing-Centric Synthesis}, which produces reasoning narratives 
    and student-like drawings conditioned on capability profiles; and 
    (3) \textbf{Diagnostic Concept Mapping}, which converts outputs into a 
    four-layer concept map linking observable features to underlying understanding 
    and suggested instructional next steps.}
    \label{fig:pipeline}
\end{figure*}

\section{DrawSim-PD Framework}

We present \textbf{DrawSim-PD}, a generative framework that simulates student-like science drawings accompanied by reasoning narratives and teacher-facing diagnostic concept maps. The framework addresses two core challenges: (1) producing synthetic artifacts that maintain both scientific validity and developmental authenticity (the ``inverse problem'' of generating specific errors), and (2) providing diagnostic scaffolding to support teacher interpretation and calibration activities.

\subsection{Framework Overview and Rationale}

As illustrated in Fig.~\ref{fig:pipeline}, DrawSim-PD comprises three integrated modules coordinated through shared \textbf{capability profiles}. The framework takes as input an NGSS performance expectation, a target grade level, and a desired performance level. It then generates:
\begin{enumerate}
    \item A \textbf{first-person reasoning narrative} simulating the student's internal monologue and scientific vocabulary.
    \item A \textbf{hand-drawn style scientific illustration} reflecting grade-appropriate motor skills and specific, realistic misconceptions.
    \item A \textbf{structured diagnostic concept map} linking visual observations to underlying understanding, serving as an answer key for teacher diagnosis.
\end{enumerate}

Simulating pedagogically valid student drawings (\emph{i.e.}, scientific modeling) necessitates solving three interconnected challenges. \emph{Challenge 1: Controllable misconceptions.} Visual misconceptions must manifest through specific spatial arrangements, missing elements, and incorrect relationships, requiring structured control rather than stochastic perturbation. \emph{Challenge 2: Cross-modal coherence.} A simulated student who ``doesn't understand cyclical processes'' must produce drawings lacking return arrows, narratives expressing confusion, and concept maps identifying this gap; independent generation risks hallucinating contradictory competencies. \emph{Challenge 3: Curriculum grounding.} Misconceptions must align with documented learning progressions, not arbitrary errors. DrawSim-PD addresses these through capability profiles as simulated cognitive states (Challenge~1), unified generation conditioned on shared profiles (Challenge~2), and automated NGSS decomposition (Challenge~3).

\subsection{NGSS-Aligned Student Simulator}

\begin{figure*}[t]
    \centering
    \includegraphics[width=\textwidth]{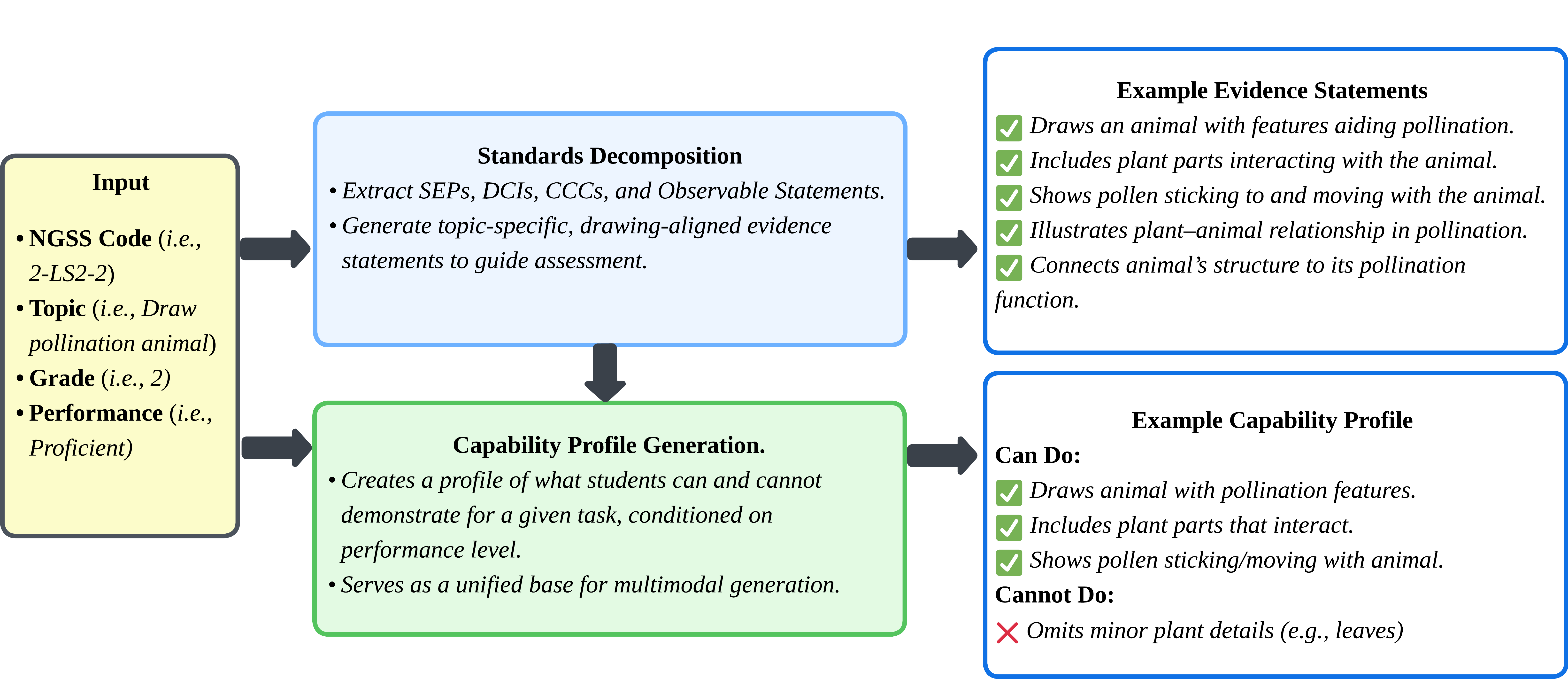} 
    \caption{\textbf{NGSS-Aligned Student Simulator.} This module converts NGSS 
    performance expectations (Science and Engineering Practices, Disciplinary 
    Core Ideas, Crosscutting Concepts) into capability profiles encoding what 
    students at each performance level (Emergent, Developing, Proficient, Advanced) 
    can and cannot demonstrate.}
    \label{fig:module1}
\end{figure*}

\paragraph{Standards Decomposition.}
As illustrated in Fig.~\ref{fig:module1}, the simulation process begins by decomposing high-level NGSS performance expectations into assessable, task-specific evidence statements. For a given NGSS code (\emph{e.g.}, K-ESS3-1: \textit{Use a model to represent the relationship between the needs of different plants or animals...}), the system retrieves the associated Science and Engineering Practices (SEPs), Disciplinary Core Ideas (DCIs), and Crosscutting Concepts (CCCs) from curated metadata.

Since NGSS standards are written for broad curriculum design rather than specific assessment tasks, we utilize GPT-4o~\cite{openai2023gpt4} to reformulate these abstract elements into 5--8 targeted \textbf{observable evidence statements} aligned with drawing-based tasks. These statements describe concrete visual features that should appear in student work (\emph{e.g.}, ``Labeling body parts'', ``Depicting habitat features''). This decomposition maintains fidelity to NGSS intentions while providing concrete, atomic constraints for the generative process.

\paragraph{Performance Level Framework.}
To represent realistic student variation, the framework employs four performance levels inspired by common rubric schemes used in K--12 science assessment~\cite{Herman2016}:
\begin{itemize}
    \item \textbf{Level 1 (Emergent):} Minimal conceptual integration; partial or inaccurate representation; often lacks labels or clear spatial organization.
    \item \textbf{Level 2 (Developing):} Basic concept recognition with limited integration; often contains specific ``hybrid'' misconceptions (mixing scientific and intuitive ideas).
    \item \textbf{Level 3 (Proficient):} Grade-appropriate reasoning with integrated three-dimensional understanding; meets the standard.
    \item \textbf{Level 4 (Advanced):} Sophisticated reasoning with accurate, complete representations; often includes details beyond the grade-level requirement.
\end{itemize}

\paragraph{Capability Profile Generation.}
For each performance level, we construct a \textbf{Capability Profile} specifying two sets of constraints:
\begin{itemize}
    \item \textbf{Can Do:} A subset of evidence statements representing concepts the student has mastered.
    \item \textbf{Cannot Yet Do:} A subset representing specific gaps, misconceptions, or skills not yet developed.
\end{itemize}
This representation functions as a shared reasoning state across all generation stages. For illustration, a \textbf{Developing-level} profile for a water cycle task might include basic evaporation understanding (\textit{Can Do:} ``Identify water rising'') while lacking condensation mechanisms (\textit{Cannot Yet Do:} ``Connect clouds to precipitation'').

\subsection{Drawing-Centric Synthesis}

\begin{figure*}[t]
    \centering
    \includegraphics[width=\textwidth]{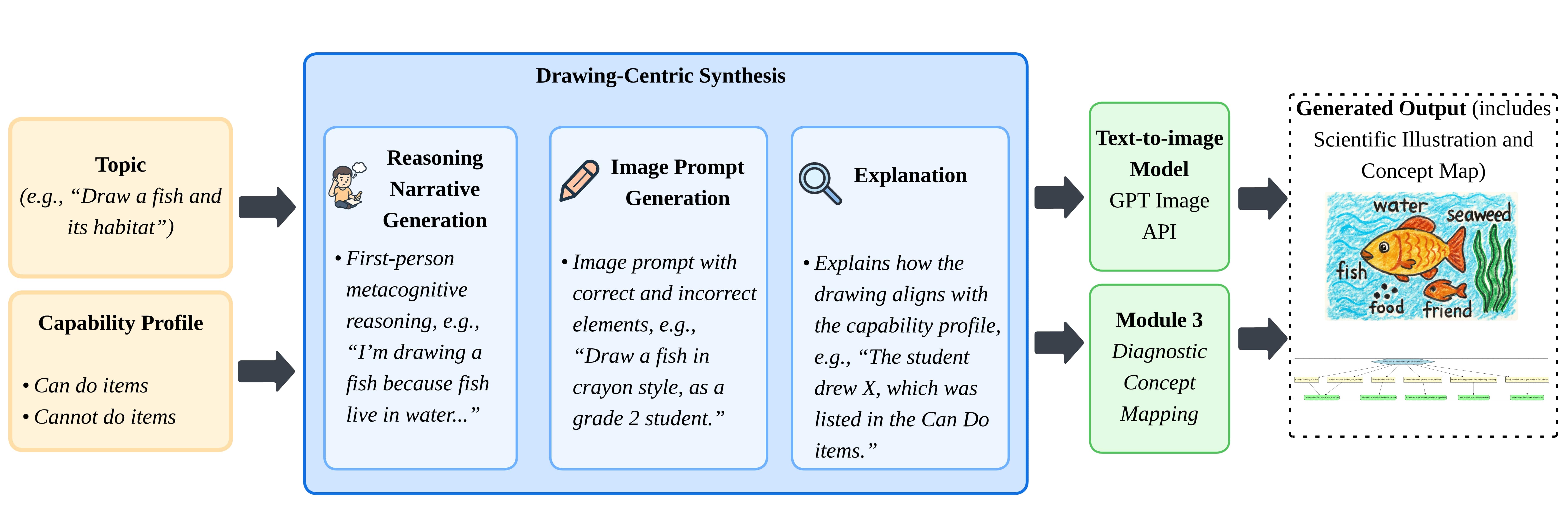} 
    \caption{\textbf{Drawing-Centric Synthesis.} This module generates student-like 
    drawings by conditioning text-to-image generation on capability profiles, 
    coordinating reasoning narratives with visual outputs to maintain coherence 
    across modalities.}
    \label{fig:module2}
\end{figure*}

\paragraph{Coordinated Multimodal Generation.}
As illustrated in Fig.~\ref{fig:module2}, the capability profile coordinates generation across modalities. We explored three generation strategies: (1)~independent generation of each modality (high risk of contradiction), (2)~sequential generation passing outputs between stages, and (3)~\textbf{unified generation}, producing all outputs in a single coordinated pass. In our iterative development, the unified approach yielded the most consistent alignment, effectively preventing ``error propagation'' where a hallucinated detail in one modality conflicts with another.

In this approach, GPT-4o generates: (1)~a first-person reasoning narrative reflecting the profile's vocabulary constraints, (2)~a detailed, grade-appropriate image prompt embedding both correct and incorrect elements, and (3)~a prompt-profile alignment explanation used for validation.

\paragraph{Reasoning Narrative Generation.}
The framework generates first-person narratives simulating student thinking during the drawing process. These narratives reflect the capability profile's constraints, expressing both understanding and limitations in developmentally appropriate language. For example: \textit{``I'm drawing the fish in the water. I know they have fins to swim, but I don't know where they sleep or what they eat.''} We use structured prompting to ensure narratives accurately reflect the specific misconceptions in the profile, rather than generic confusion.

\paragraph{Scientific Illustration Generation.}
Visual generation uses the reasoning narrative and capability profile to create grade-appropriate image prompts. This is a non-trivial ``inverse generation'' task: standard models are optimized to be helpful and correct, so they often resist generating ``bad'' diagrams. To overcome this, the system composes prompts specifying:
\begin{enumerate}
    \item \textbf{Positive Constraints:} Elements from the \textit{Can Do} set (\emph{e.g.}, ``Draw a sun and arrows pointing up'').
    \item \textbf{Negative Constraints:} Omissions or distortions reflecting \textit{Cannot Yet Do} aspects (\emph{e.g.}, ``Do NOT include clouds or rain; do NOT connect the cycle back to the ground'').
    \item \textbf{Stylistic Constraints:} Developmental markers such as ``hand-drawn crayon style,'' ``uneven lines,'' or ``simple 2D perspective'' appropriate for the target grade.
\end{enumerate}

After evaluating multiple text-to-image models including Stable Diffusion XL~\cite{podell2023sdxl} and FLUX~\cite{BlackForestLabs2025}, we utilize the OpenAI GPT Image 1 API~\cite{openai2025imagegen}, which offered superior instruction following for negative constraints and consistently adhered to the requested ``imperfect'' artistic styles.

\subsection{Diagnostic Concept Mapping}

\begin{figure*}[t]
    \centering
    \includegraphics[width=\textwidth]{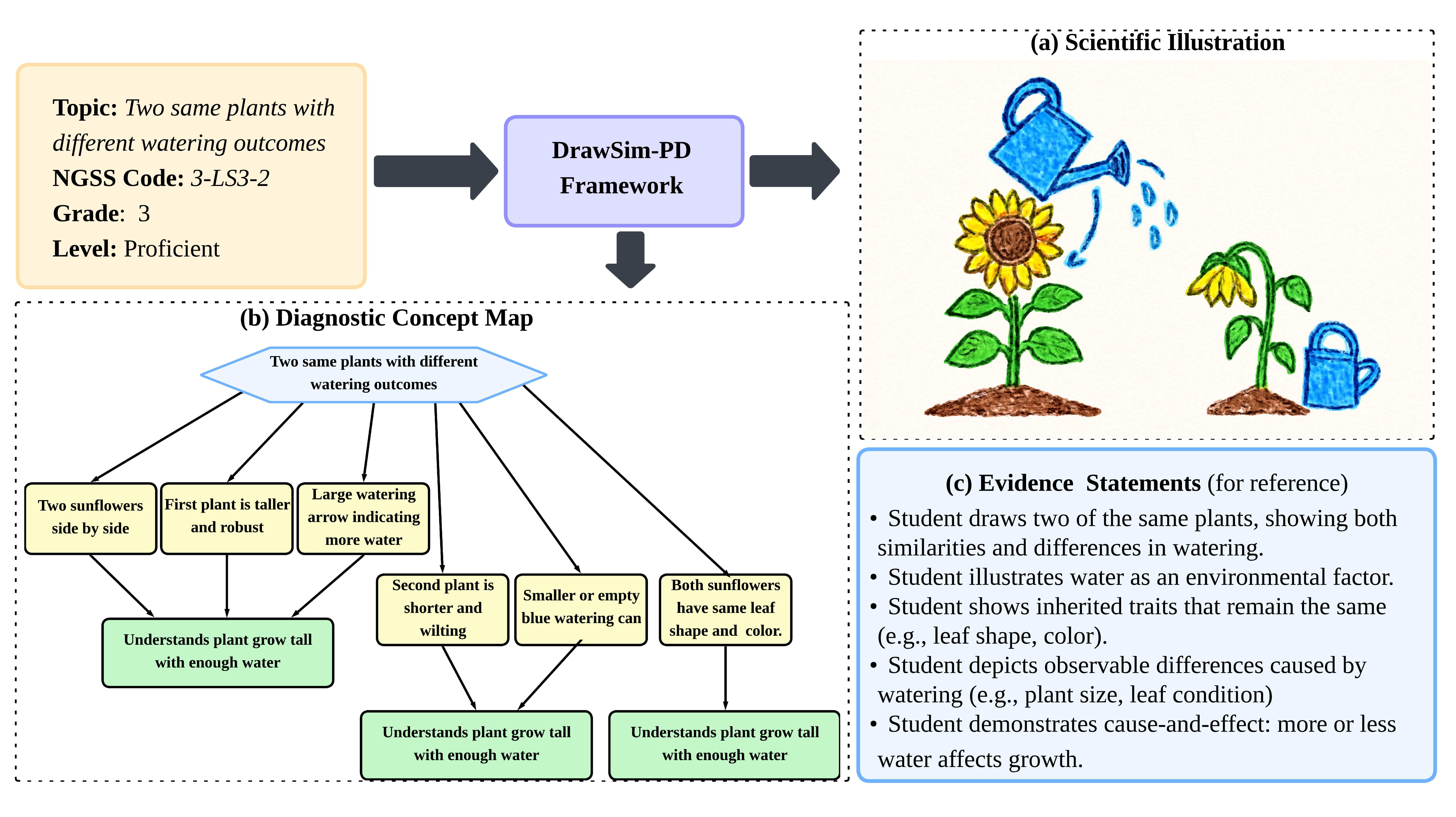}
    \caption{\textbf{Example DrawSim-PD output.} Given topic, NGSS code, grade, 
    and performance level, the framework generates: (a)~a student-like scientific 
    illustration, (b)~a diagnostic concept map linking visual features to 
    underlying reasoning, and (c)~reference evidence statements defining 
    targeted learning goals.}
    \label{fig:example}
\end{figure*}

\paragraph{Teacher-Facing Diagnostic Layer.}
To support teacher interpretation and calibration activities, we transform each drawing specification into a structured diagnostic concept map. By design, \textbf{we generate concept maps from the drawing specification (prompt + profile) rather than from the rendered image itself.} This architectural choice ensures the diagnostic layer reflects the intended student understanding encoded in the capability profile, avoiding potential hallucination from visual interpretation (\emph{e.g.}, a Vision-Language Model misinterpreting a messy sketch).

\paragraph{Hierarchical Concept Map Structure.}
We adopt a four-layer representation designed for rapid teacher interpretation (see Fig.~\ref{fig:example}):
\begin{enumerate}
    \item \textbf{Topic Layer:} Single node naming the NGSS-aligned task.
    \item \textbf{Observation Layer:} Nodes describing concrete features visible in the drawing (\emph{e.g.}, ``Arrows point up'').
    \item \textbf{Understanding Layer:} Nodes interpreting these features as evidence of understanding or misconception (\emph{e.g.}, ``Understands Evaporation'').
    \item \textbf{Feedback Layer:} Nodes providing suggested instructional next steps, attached specifically to misconceptions.
\end{enumerate}
Directed edges capture hierarchical relationships: Topic $\rightarrow$ Observation $\rightarrow$ Understanding $\rightarrow$ Feedback. GPT-4o generates these maps as JSON objects, which undergo structural validation before rendering via PyGraphviz.

\subsection{Corpus Construction}

\begin{figure*}[t]
    \centering
    \includegraphics[width=\textwidth]{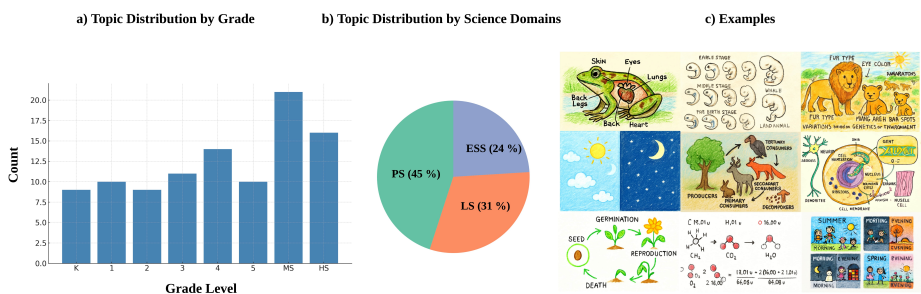} 
    \caption{\textbf{DrawSim-PD corpus composition.} (a)~Grade distribution 
    showing coverage across K--12. (b)~Domain distribution: Physical Sciences 
    (45\%), Life Sciences (31\%), Earth \& Space Sciences (24\%). 
    (c)~Representative drawings illustrating performance-level variation.}
    \label{fig:distribution}
\end{figure*}

\paragraph{Topic Curation.}
We obtained 191 NGSS performance expectations from official documentation~\cite{NGSS}. Three science education faculty reviewed these to select 100 topics where visual modeling is pedagogically central (\emph{e.g.}, systems, cycles, forces), excluding topics better suited for pure text or calculation. The resulting set spans Physical Sciences (45\%), Life Sciences (31\%), and Earth \& Space Sciences (24\%) across K--12 grade bands, as illustrated in Fig.~\ref{fig:distribution}.

\paragraph{Structured Sampling Rationale.}
For each topic, we generate 25 synthetic exemplars at each of four performance levels, yielding a total corpus of \textbf{10,000 artifacts} with structured metadata. This systematic sampling (100 topics $\times$ 4 levels $\times$ 25 exemplars) supports distinct professional development use cases:
\begin{itemize}
    \item \textbf{Calibration Exercises:} Comparing interpretations of the same topic across performance levels.
    \item \textbf{Vertical Alignment:} Comparing how a concept (\emph{e.g.}, gravity) is modeled in Grade 2 vs. Grade 5.
    \item \textbf{Misconception Analysis:} Examining a library of specific errors (\emph{e.g.}, ``broken cycles'') decoupled from confounding variables like student handwriting quality.
\end{itemize}
To our knowledge, this represents the largest collection of curriculum-aligned student drawing simulations to date. We release the full corpus, generation code, and metadata to support further research.

%% file: sec_arXiv/exp_0201.tex
\section{Experiments}

Having introduced a generation paradigm that prioritizes pedagogically meaningful 
imperfection over aesthetic accuracy, we conducted an expert feasibility study 
addressing a critical question: \emph{Can generative AI produce student-like 
science drawings that experienced educators recognize as pedagogically authentic 
and diagnostically useful?} Our evaluation provides initial evidence that 
DrawSim-PD crosses this plausibility threshold while identifying refinement 
opportunities. We address three research questions:
\vspace{-2mm}
\begin{itemize}
    \item \textbf{RQ1}: To what extent are generated artifacts aligned with 
    NGSS performance expectations and plausible for target grade bands?
    \item \textbf{RQ2}: How do capability profiles affect performance-level 
    differentiation and cross-modal consistency?
    \item \textbf{RQ3}: How do teachers perceive the usefulness of these 
    artifacts for diagnostic reasoning and professional learning?
\end{itemize}
\vspace{-2mm}
\subsection{Study Design}

Following established practices in educational technology validation~\cite{dai2024assessing,lin2025automatic}, 
we employed expert review to assess the pedagogical fidelity of the generated 
corpus. This approach is particularly appropriate for evaluating three-dimensional 
learning, which requires specialized knowledge of integrating DCIs, SEPs, and 
CCCs~\cite{fick2018does,kang2018exploring}. Our design prioritized breadth of 
coverage—collecting 480 discrete evaluations across 100 topics—over depth of 
inter-rater agreement on a small subset. This strategy aligns with the goals of 
a feasibility assessment: identifying systematic performance patterns across the 
corpus rather than establishing psychometric precision for a single item.

\Paragraph{Participants}
We recruited six subject matter experts: former or current K--12 science teachers 
currently enrolled in graduate science education programs (M = 2.75 years 
classroom teaching experience, SD = 1.84; three in-service, three former). 
Participants reported moderate-to-high familiarity with NGSS (M = 3.33/5). 
Their dual experience as practitioners and researchers provided diverse 
perspectives on both classroom plausibility and theoretical alignment.

\Paragraph{Materials and Procedure}
Each participant evaluated 80 unique DrawSim-PD artifacts through an online 
survey over one week. Artifacts were stratified by random sampling to ensure 
balanced coverage across: (1)~four performance levels (Emergent, Developing, 
Proficient, Advanced), (2)~three NGSS domains (Physical, Life, Earth \& Space), 
and (3)~four grade spans (K--2, 3--5, 6--8, 9--12). Participants viewed the 
complete output (drawing, narrative, concept map) before completing the 
instrument.

\Paragraph{Evaluation Instrument}
We designed a mixed-response instrument (Fig.~\ref{fig:instrument}) combining 
categorical judgments with 5-point Likert ratings. Items Q1--Q5 assessed strict 
NGSS alignment and performance-level matching, while Q6--Q8 assessed holistic 
plausibility and component quality. Open-ended prompts captured qualitative 
feedback.

\begin{figure}[t]
\centering
\begin{tcolorbox}[
  enhanced,
  colback=white,
  colframe=gray,
  boxrule=0.6pt,
  arc=1.2mm,
  left=6pt,right=6pt,top=6pt,bottom=6pt,
  width=\columnwidth
]
\small
\setlist[enumerate]{leftmargin=*, itemsep=0.2em, topsep=0.2em, parsep=0pt}

\begin{tcolorbox}[colback=gray!10, colframe=gray!10, boxrule=0pt, arc=1mm, left=4pt,right=4pt,top=2pt,bottom=2pt]
\textbf{NGSS Alignment} \hfill \emph{(Yes / Partially / No)}
\end{tcolorbox}
\begin{enumerate}[label=\textbf{Q\arabic*:}]
  \item Does the topic align with the NGSS Performance Expectation?
  \item Does the drawing represent the disciplinary core ideas?
  \item Does the drawing align logically with the given prompt?
  \item Does the drawing align with the capability statements?
  \item Does the drawing match the assigned performance level?
\end{enumerate}

\begin{tcolorbox}[colback=gray!10, colframe=gray!10, boxrule=0pt, arc=1mm, left=4pt,right=4pt,top=2pt,bottom=2pt]
\textbf{Grade-Band Plausibility} \hfill \emph{(Yes / Partially / No)}
\end{tcolorbox}
\begin{enumerate}[label=\textbf{Q\arabic*:}, resume]
  \item Does the drawing appear plausible for the target grade band?
\end{enumerate}

\begin{tcolorbox}[colback=gray!10, colframe=gray!10, boxrule=0pt, arc=1mm, left=4pt,right=4pt,top=2pt,bottom=2pt]
\textbf{Component Quality} \hfill \emph{(1--5 Likert)}
\end{tcolorbox}
\begin{enumerate}[label=\textbf{Q\arabic*:}, resume]
  \item Does the concept map represent the reasoning in the drawing?
  \item Does the work maintain plausible scientific relationships for the level?
\end{enumerate}

\end{tcolorbox}
\vspace{-2mm}
\caption{\textbf{Expert Evaluation Instrument.} The mixed-method survey used for the feasibility study, assessing strict NGSS alignment (Q1--Q5), developmental plausibility (Q6), and pedagogical utility (Q7--Q8).}
\label{fig:instrument}
\vspace{-4mm}
\end{figure}

\subsection{Results: NGSS Alignment and Plausibility} \label{sec:results}

Participants evaluated 480 unique artifacts, providing systematic coverage of 
the corpus. Results exceeded expectations for a first-generation system 
(Tab.~\ref{tab:results}).

\begin{table}[t]
\centering
\caption{Expert Evaluation Results (N=480 evaluations).}
\label{tab:results}
\resizebox{0.9\linewidth}{!}{
\begin{tabular}{lccc|ccccc}
\toprule
\multirow{2}{*}{\textbf{Dimension} }&  \multicolumn{3}{c}{\textbf{Binary Judgment (\%)}} & \multicolumn{5}{c}{\textbf{Likert Rating (\%)}} \\
\cmidrule(lr){2-4}\cmidrule(lr){5-9}
&\textbf{Yes} & \textbf{Partially} & \textbf{No } & \textbf{1} & \textbf{2} & \textbf{3} & \textbf{4} & \textbf{5}  \\
\midrule
\rowcolor{gray!10}\multicolumn{9}{@{}l}{\textit{NGSS Alignment (Q1--Q5)}} \\
Q1: Topic-PE Alignment & 89.58 & 8.75 & 1.67 & -- & -- & -- & -- & -- \\
Q2: DCI Representation & 84.17 & 13.75 & 2.08 & -- & -- & -- & -- & -- \\
Q3: Drawing-Prompt Coherence & 86.66 & 12.92 & 0.42 & -- & -- & -- & -- & -- \\
Q4: Capability Statement Match & 75.00 & 24.17 & 0.83 & -- & -- & -- & -- & --  \\
Q5: Performance Level Match & 73.75 & 19.17 & 7.08 & -- & -- & -- & -- & --  \\
\midrule
\rowcolor{gray!10}\multicolumn{9}{@{}l}{\textit{Grade-Band Plausibility (Q6)}} \\
Q6: Grade-Level Authenticity & 60.42 & 32.92 & 6.66 & -- & -- & -- & -- & -- \\
\midrule
\rowcolor{gray!10}\multicolumn{9}{@{}l}{\textit{Component Quality (Q7--Q8)}} \\
Q7: Concept Map Quality & -- & -- & -- & 0.0 & 3.3 & 19.2 & 47.9 & 29.6  \\
Q8: Scientific Accuracy & -- & -- & -- & 5.0 & 13.3 & 19.6 & 32.1 & 30.0  \\
\bottomrule
\end{tabular}
}
\vspace{-3mm}
\end{table}

\Paragraph{NGSS Alignment}
Artifacts demonstrated strong alignment with NGSS standards. Topic-PE 
alignment achieved 89.6\% full agreement (Q1), indicating successful mapping 
of standards to drawing tasks. Disciplinary Core Idea representation 
(Q2: 84.2\%) and drawing-prompt coherence (Q3: 86.7\%) confirmed that 
generated content captures required scientific concepts. Notably, explicit 
disagreements (``No'') remained below 2.1\% across these core alignment items.

\Paragraph{Performance Differentiation}
The alignment of artifacts with capability statements (Q4: 75.0\% Yes) and 
performance levels (Q5: 73.8\% Yes) suggests that the capability profiles 
effectively control the sophistication of the output. Combined positive 
responses (Yes + Partially) exceeded 92\% for both items. The prevalence of 
``Partially'' ratings (approx. 20\%) likely reflects the inherent ambiguity 
in boundary cases between adjacent performance levels—a known challenge in 
human assessment rubrics.

\Paragraph{Grade-Band Plausibility}
%
Grade-band plausibility (Q6) exceeded expectations for simulating developmental 
characteristics: 93.3\% of artifacts were rated as plausible or partially 
plausible. Given the difficulty capturing age-specific motor skills and 
aesthetic choices, this suggests the system encodes key 
developmental constraints. As detailed in Sec.~\ref{sec:qualitative}, 
plausibility was highest for middle grades (3--8), with slight degradation at 
extremes (K--2 and 9--12).

\Paragraph{Component Quality}
Diagnostic concept maps (Q7) received favorable ratings (77.5\% rated 4 or 5), 
indicating they accurately externalized the reasoning embedded in the drawings. 
Scientific plausibility (Q8) showed higher variance (18.3\% rated 1 or 2). This 
reflects a deliberate design tension: the system \emph{intentionally} generates 
incorrect elements to simulate misconceptions. Some evaluators may have 
penalized these intended errors as scientific inaccuracies rather than 
pedagogical features.

\subsection{Results: Cross-Modal Consistency} \label{sec:cross_modal}

To complement expert judgments, we examined semantic alignment between 
generated components using CLIP similarity scores across 1,200 sampled 
artifacts. We treat CLIP similarity as an engineering consistency diagnostic 
rather than a proxy for educational validity. For reference, CLIP similarity 
for unrelated image-text pairs typically falls below 0.15.

As shown in Tab.~\ref{tab:coherence}, Concept Map-Drawing alignment achieved 
the highest consistency (0.606), suggesting effective capture of visual content 
in the structured maps. Text-Drawing consistency decreased slightly with grade 
level (0.364 to 0.348) and performance level (0.362 to 0.349). Rather than 
indicating system degradation, this pattern likely reflects a fundamental 
characteristic of science assessment: as reasoning becomes more advanced 
(Level 4) and abstract (Grades 9--12), it becomes increasingly difficult to 
represent via static visual depictions.

\begin{table}[t]
\centering
\caption{Cross-modal consistency (CLIP similarity, N = 1,200). Overall score 
is the mean of three pairwise comparisons.}
\label{tab:coherence}
\resizebox{0.7\linewidth}{!}{
\begin{tabular}{lcccc}
\toprule
& \textbf{Text--Draw} & \textbf{CMap--Draw} & \textbf{Text--CMap} & \textbf{Overall} \\
\midrule
Overall & 0.356 & 0.606 & 0.273 & 0.412 \\
\midrule
\rowcolor{gray!10}\multicolumn{5}{@{}l}{\textit{By Level}} \\
\quad Emergent & 0.362 & 0.589 & 0.250 & 0.400 \\
\quad Developing & 0.360 & 0.607 & 0.266 & 0.411 \\
\quad Proficient & 0.354 & 0.614 & 0.280 & 0.416 \\
\quad Advanced & 0.349 & 0.614 & 0.293 & 0.419 \\
\midrule
\rowcolor{gray!10}\multicolumn{5}{@{}l}{\textit{By Grade}} \\
\quad K--2 & 0.364 & 0.583 & 0.263 & 0.403 \\
\quad 9--12 & 0.348 & 0.629 & 0.286 & 0.421 \\
\bottomrule
\end{tabular}
}
\vspace{-3mm}
\end{table}

\subsection{Results: Role of Capability Profiles (Ablation Study)}

We investigated the necessity of capability profiles by generating outputs 
with and without profile conditioning (160 artifacts: 20 topics $\times$ 4 levels 
$\times$ 2 conditions). Without capability profiles, the system produced 
uniformly detailed drawings regardless of the assigned performance level, 
failing to capture developmental constraints.

\begin{figure*}[t]
  \centering
  \setlength{\tabcolsep}{4pt}
  \renewcommand{\arraystretch}{1.2}
  \resizebox{\linewidth}{!}{
  \begin{tabular}{c c c c}
    & \textbf{Cell under Microscope} & \textbf{Food Chain} & \textbf{Plant Lifecycle} \\
    \textbf{Level 1} &
    \subcaptionbox{}[.30\linewidth]{%
      \begin{minipage}{\linewidth}
        \includegraphics[width=0.49\linewidth]{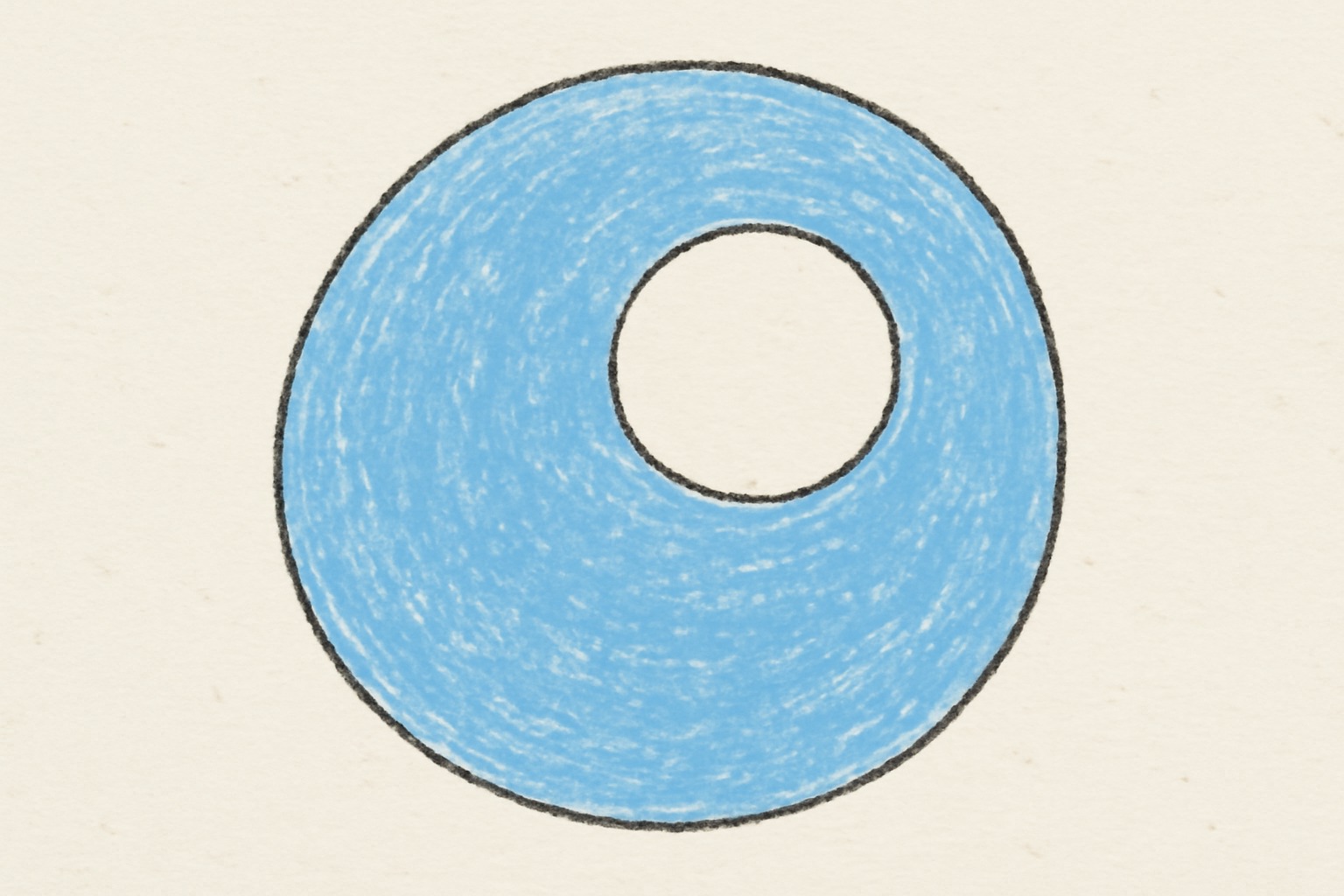}%
        \includegraphics[width=0.49\linewidth]{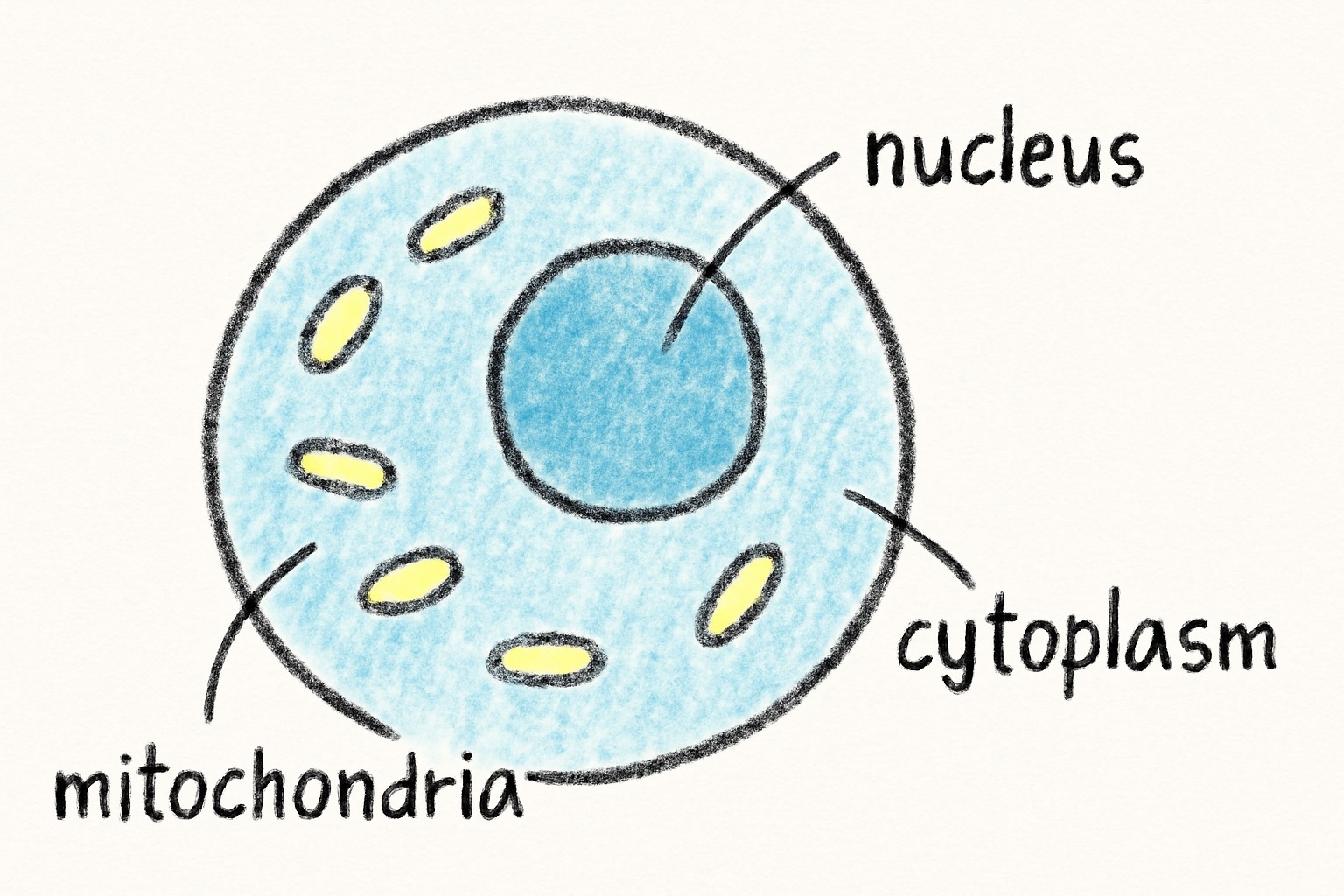}\\
        \centering \scriptsize (With Profile \quad Without Profile)
      \end{minipage}} &
    \subcaptionbox{}[.30\linewidth]{%
      \begin{minipage}{\linewidth}
        \includegraphics[width=0.49\linewidth]{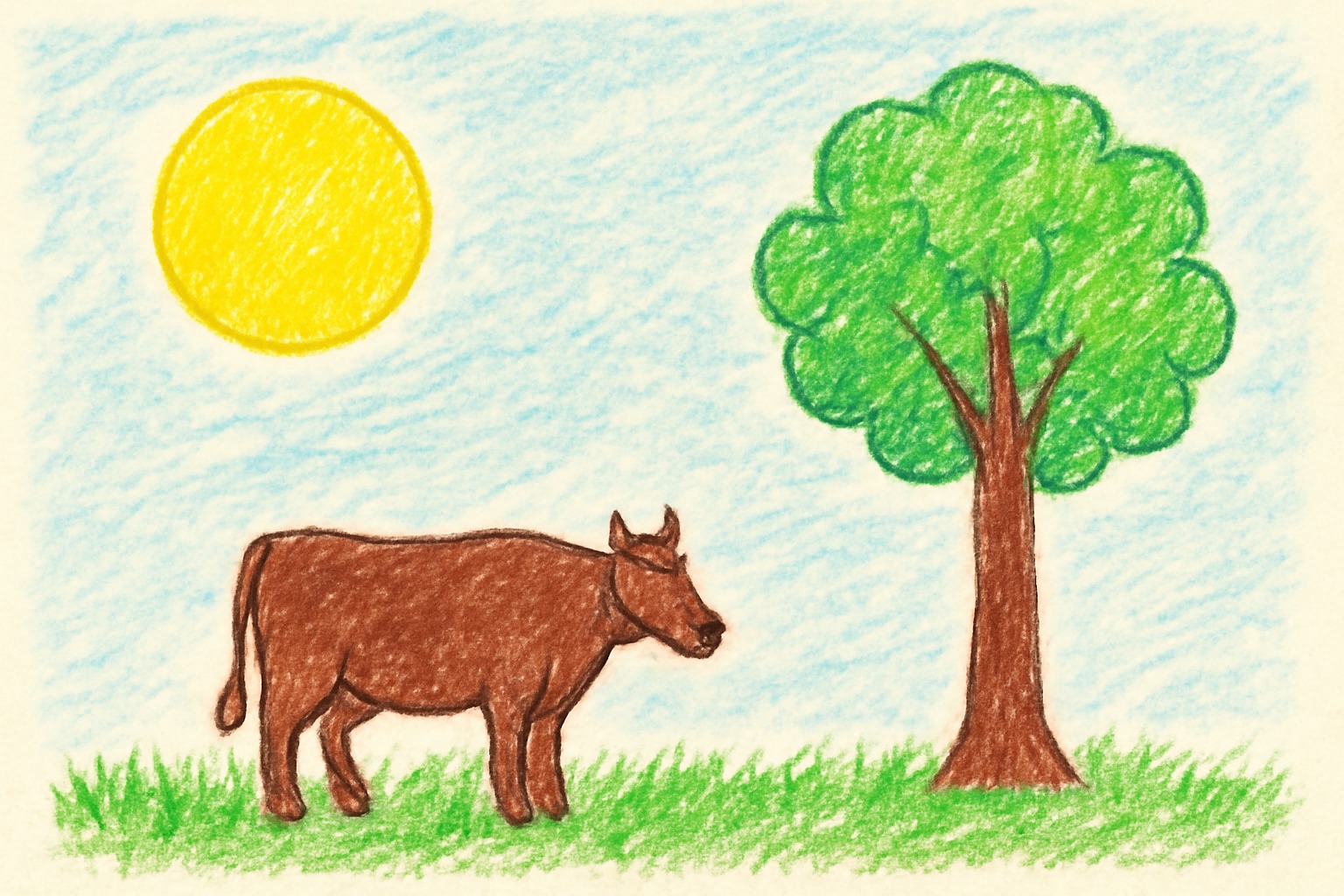}%
        \includegraphics[width=0.49\linewidth]{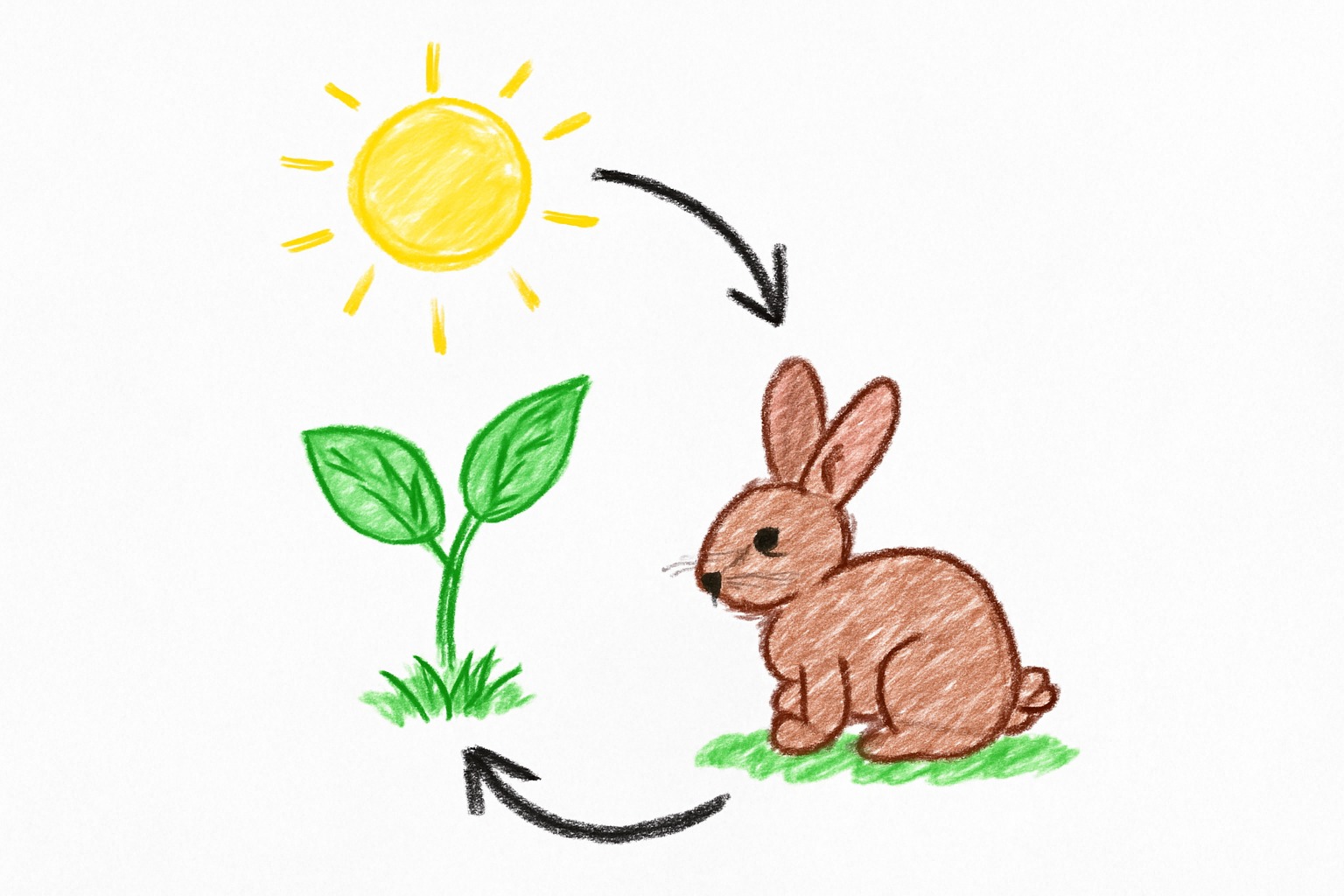}\\
        \centering \scriptsize (With Profile \quad Without Profile)
      \end{minipage}} &
    \subcaptionbox{}[.30\linewidth]{%
      \begin{minipage}{\linewidth}
        \includegraphics[width=0.49\linewidth]{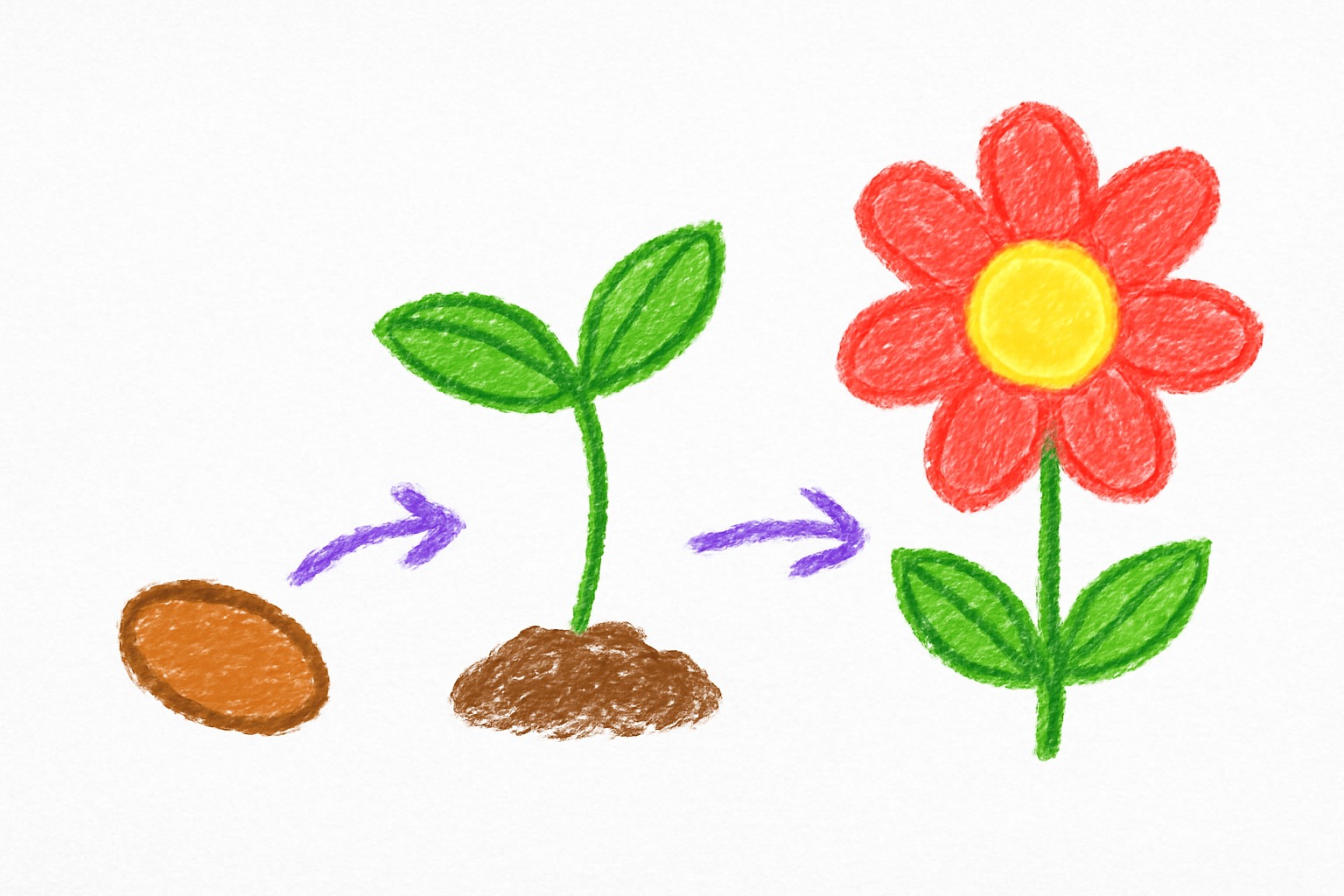}%
        \includegraphics[width=0.49\linewidth]{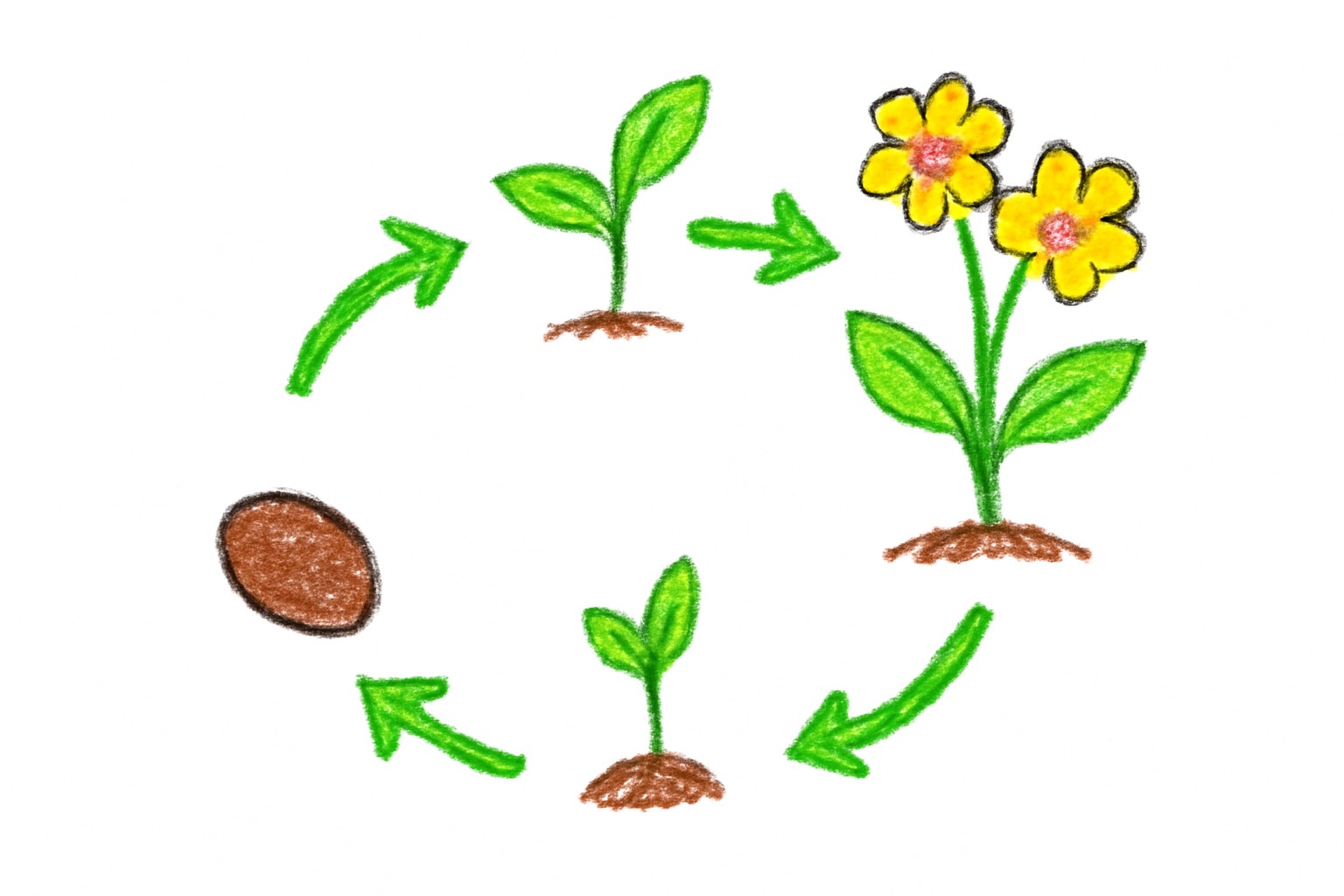}\\
        \centering \scriptsize (With Profile \quad Without Profile)
      \end{minipage}} \\
    \textbf{Level 4} &
    \subcaptionbox{}[.30\linewidth]{%
      \begin{minipage}{\linewidth}
        \includegraphics[width=0.49\linewidth]{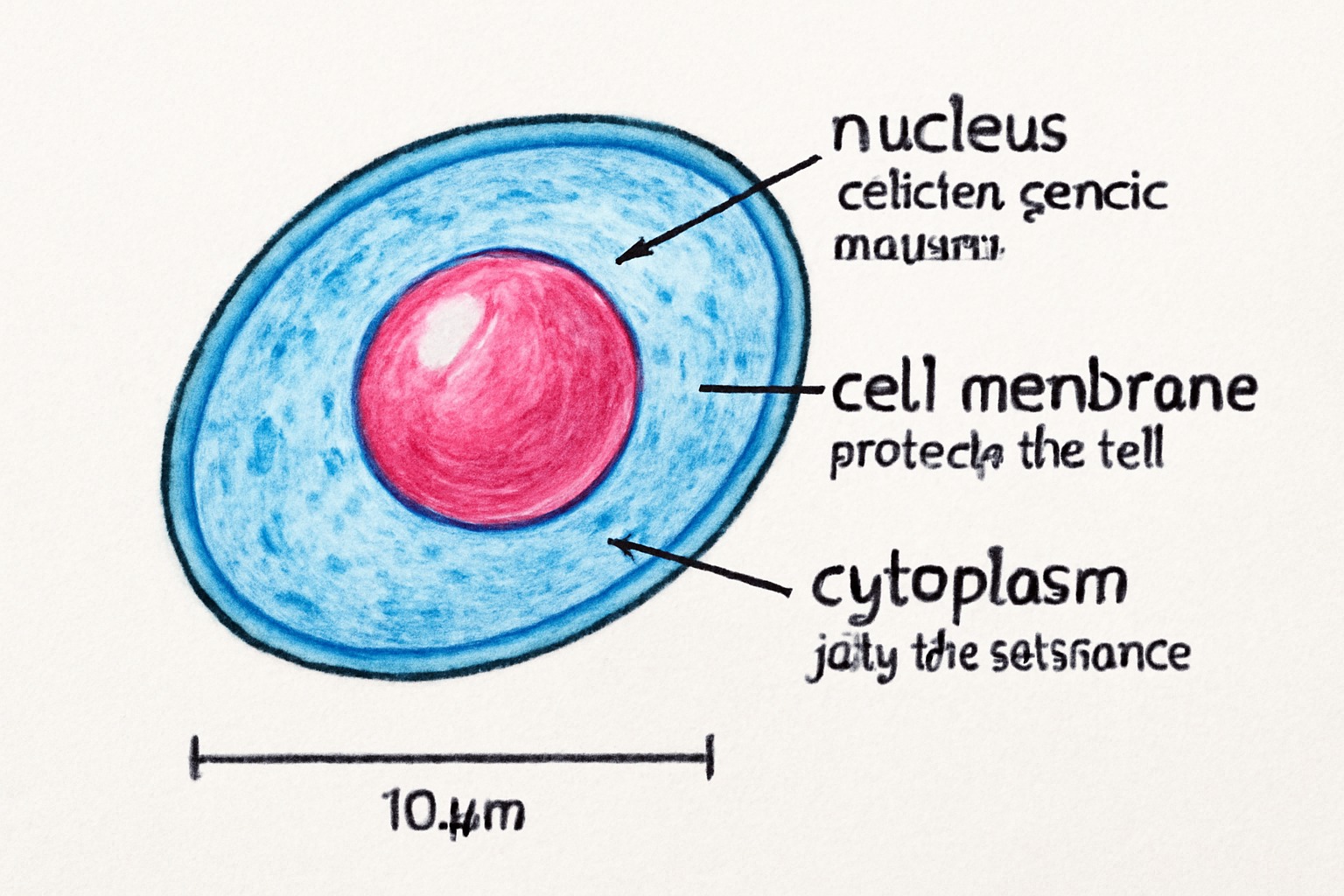}%
        \includegraphics[width=0.49\linewidth]{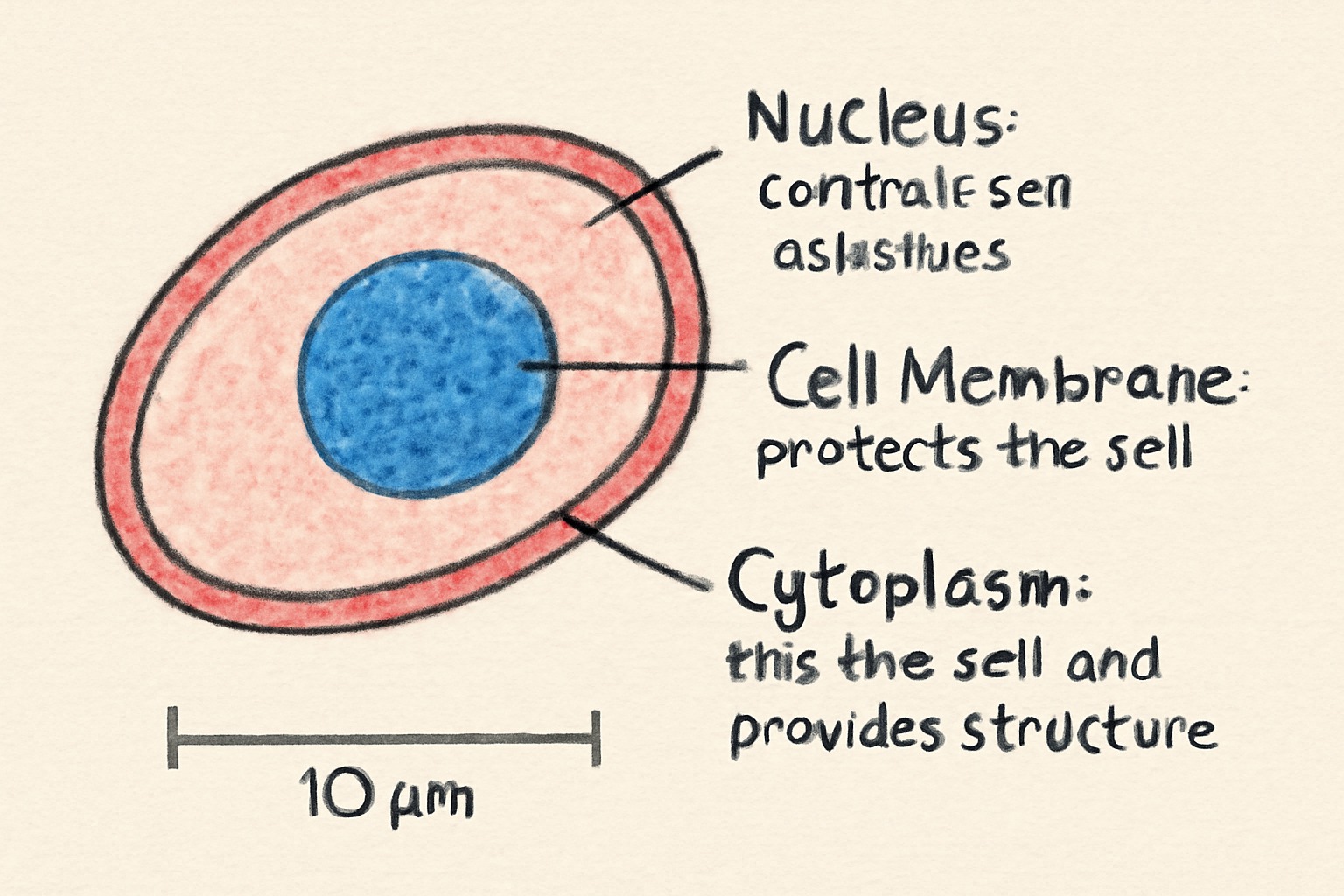}\\
        \centering \scriptsize (With Profile \quad Without Profile)
      \end{minipage}} &
    \subcaptionbox{}[.30\linewidth]{%
      \begin{minipage}{\linewidth}
        \includegraphics[width=0.49\linewidth]{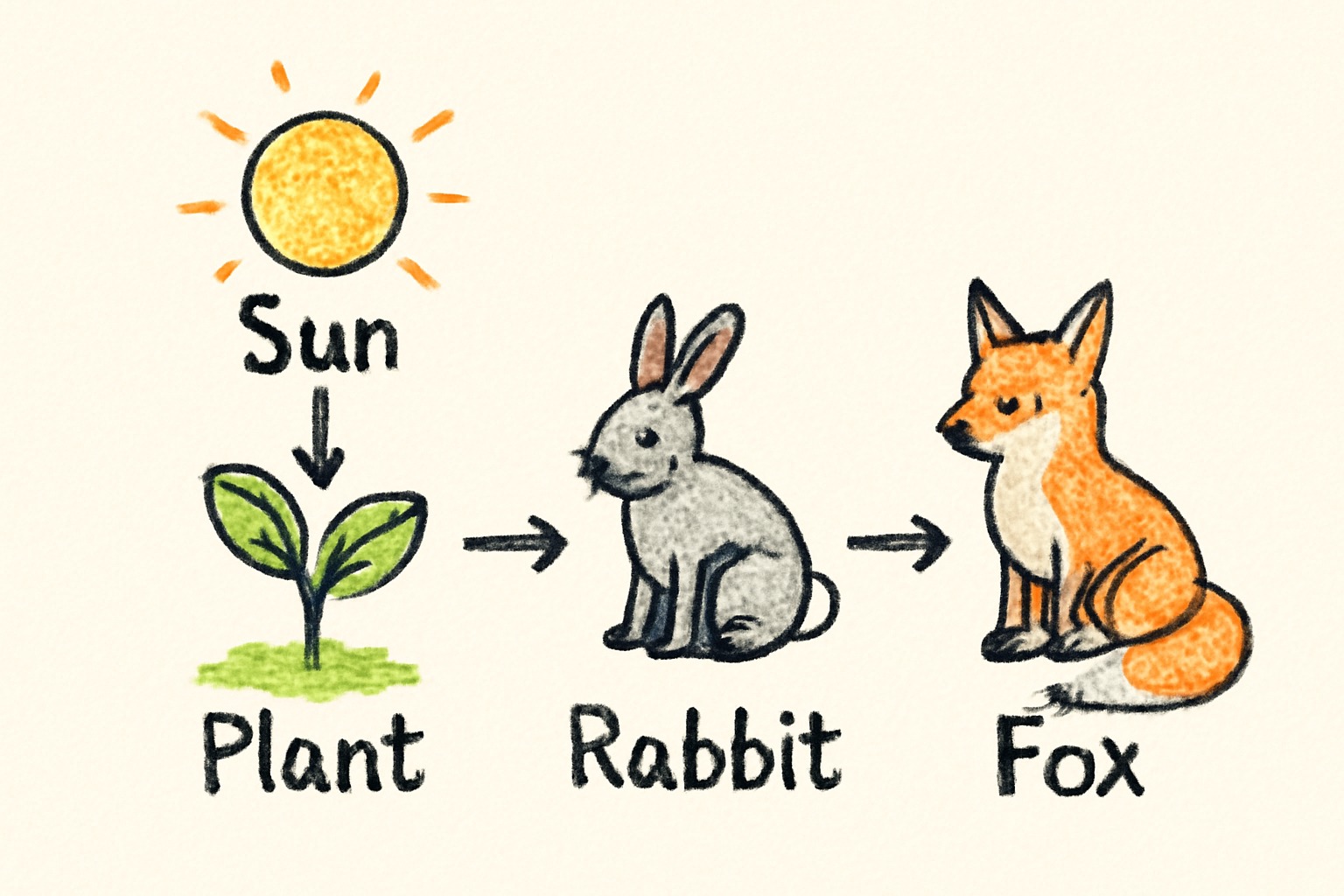}%
        \includegraphics[width=0.49\linewidth]{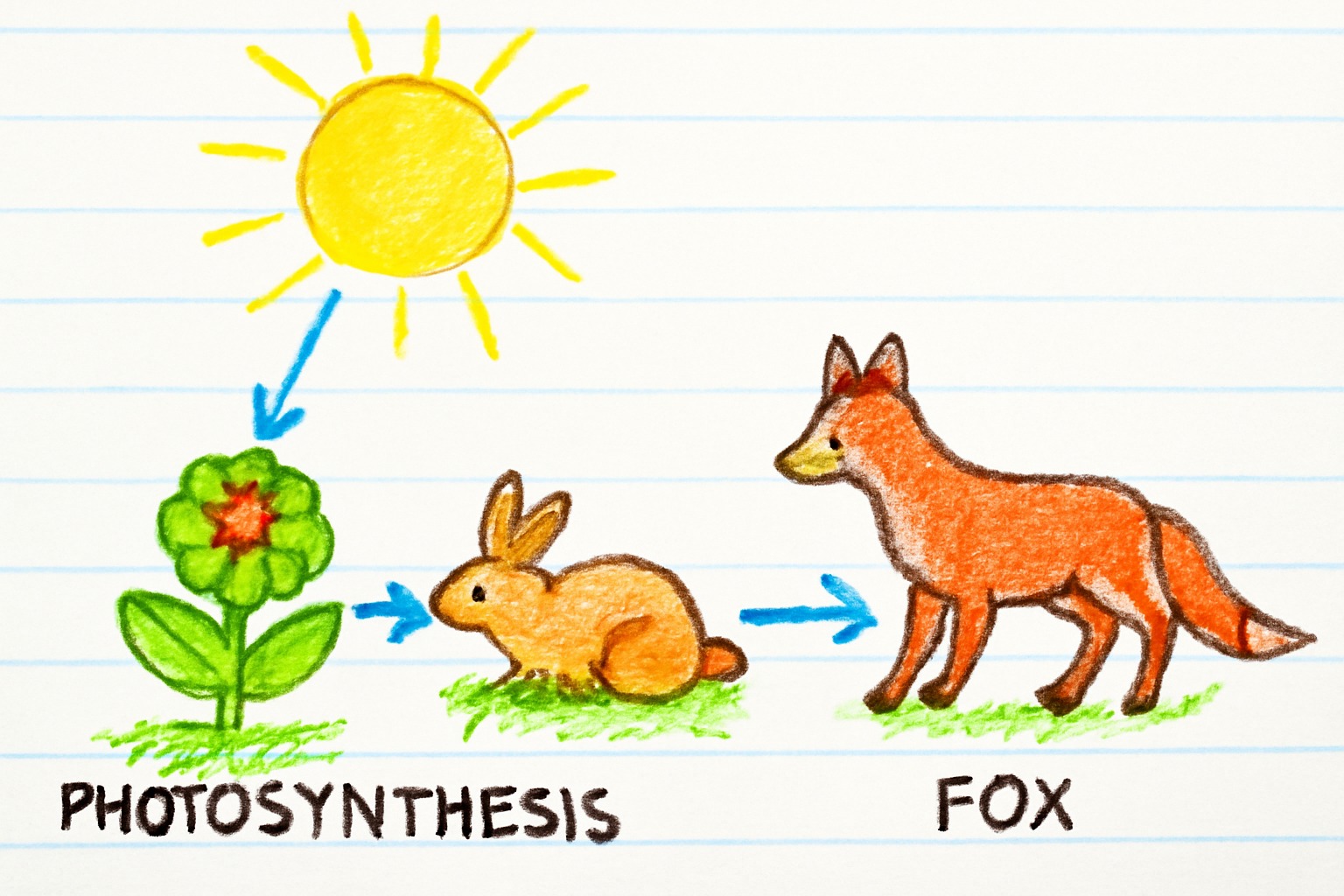}\\
        \centering \scriptsize (With Profile \quad Without Profile)
      \end{minipage}} &
    \subcaptionbox{}[.30\linewidth]{%
      \begin{minipage}{\linewidth}
        \includegraphics[width=0.49\linewidth]{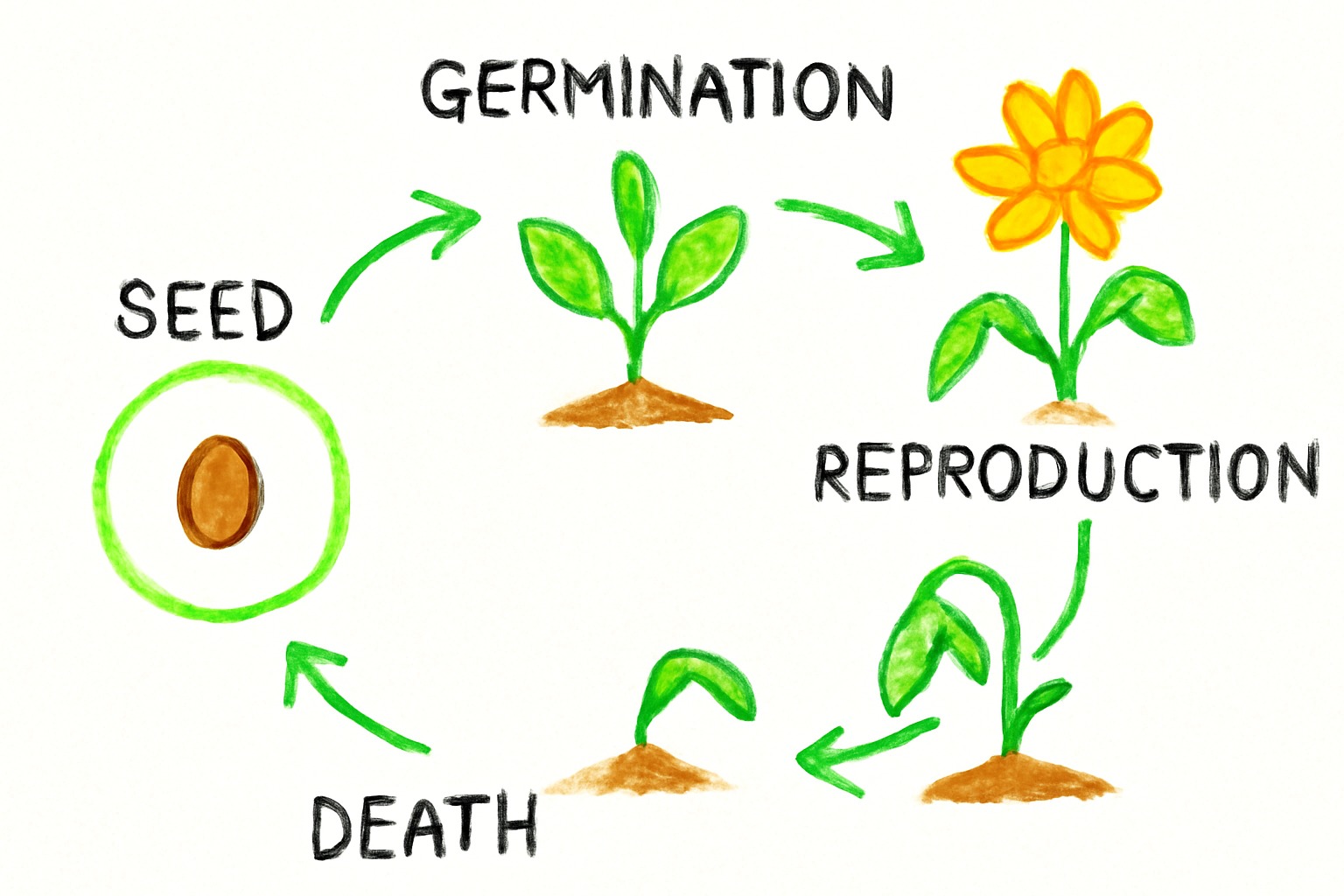}%
        \includegraphics[width=0.49\linewidth]{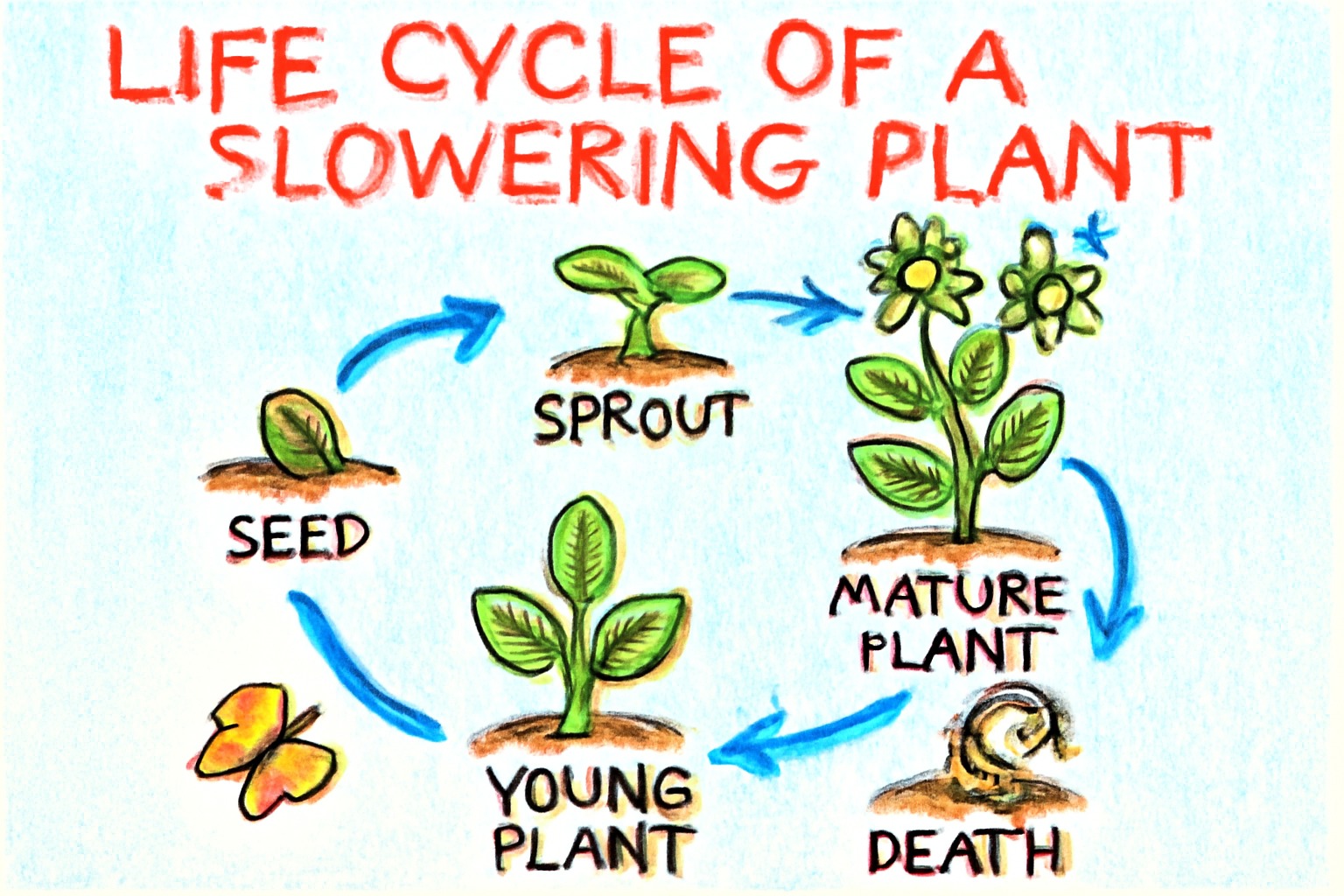}\\
        \centering \scriptsize (With Profile \quad Without Profile)
      \end{minipage}} \\
  \end{tabular}
  }
\vspace{-2mm}
\caption{\textbf{Ablation Study: The Role of Capability Profiles.} Comparisons of outputs generated \textbf{with profiles} (left) versus \textbf{without profiles} (right). With profiles, the system correctly differentiates between Level 1 (simplified, partial) and Level 4 (complete). Without profiles, the system reverts to a generic "average" complexity, failing to simulate developmental progression. Quantitative metrics confirm this collapse in variance (SD 0.31 vs. 0.08).}
\label{fig:component-grid}
\vspace{-4mm}
\end{figure*}

To quantify this, we computed the standard deviation of image complexity 
(Canny edge density normalized by area) across the four performance levels 
within each topic. \textbf{With profiles}, cross-level complexity variance was 
substantial (mean SD = 0.31), reflecting appropriate differentiation. 
\textbf{Without profiles}, variance collapsed (mean SD = 0.08), indicating 
static output complexity. This nearly fourfold difference demonstrates that 
capability profiles are essential for achieving performance-level 
differentiation. Fig.~\ref{fig:component-grid} illustrates this effect.

\subsection{Teacher Perspectives on Utility} \label{sec:qualitative}

We conducted a qualitative thematic analysis using inductive techniques~\cite{Fereday2006} to examine how expert participants perceived the generated corpus. To ensure rigor, two authors collaboratively coded the open-ended responses and refined categories through iterative discussion, following established guidelines for reliability in qualitative research~\cite{mcdonald_reliability_2019}. Five key themes emerged regarding system utility, authenticity, and design implications.

\Paragraph{Plausibility Crossed the Authenticity Threshold}
Participants consistently described outputs as immediately recognizable as student-like due to their simplicity and visual style. P6 noted, \textit{``This looks like the work of a real student,''} expressing confidence that the system could reflect classroom realities. However, authenticity was sometimes compromised by an "uncanny valley" of neatness: P1 explained, \textit{``The composition is too clear... the student's arrangement would be more chaotic.''} Participants stressed that true authenticity requires balancing scientific errors with stylistic imperfections, noting that \textit{``when the drawing skill level matches the grade bands, it is more like real student thinking''} (P5).

\Paragraph{Grade-level Differentiation Succeeded for Core Grades}
The system's ability to stratify low- and high-level work was seen as a major strength, particularly in the middle grades (3--8). Discrepancies concentrated at the developmental extremes, highlighting specific challenges for prompt engineering. P6 observed occasional underestimations of lower-grade students, while P5 noted that \textit{``kindergarten drawings were above students' level''} regarding motor control, whereas high school examples sometimes relied on scientific errors \textit{``too obvious to be considered''} for that level. This suggests that capability profiles need tighter constraints for K--2 motor skills and 9--12 abstract reasoning.

\Paragraph{Misconceptions were Pedagogically Relevant but "Mechanical"}
Artifacts successfully reflected classroom-like misunderstandings, such as superficial reasoning or missing causal links, which teachers found useful for illustrating common pitfalls. However, a subtle distinction emerged regarding the \emph{nature} of these errors. P2 remarked, \textit{``AI's misconceptions appear mechanical and rigid... students' misconceptions are slightly more flexible.''} This finding is critical for AIED researchers: while the system effectively captures \emph{what} students get wrong (content), it may not yet fully capture \emph{how} they get it wrong (the fluid, often tentative nature of student thinking).

\Paragraph{Cross-modal Integration Enhanced Credibility}
Teachers valued the coordinated presentation of drawings, narratives, and concept maps, reporting that this multimodal structure made the outputs more credible and easier to use as teaching artifacts. P4 described the components as \textit{``generally consistent,''} and P6 found \textit{``no significant discrepancies.''} Occasional mismatches—such as concept maps containing vocabulary not present in the drawing—were noted, but the consensus was that the diagnostic map successfully externalized the reasoning implied by the drawing.

\Paragraph{Utility depends on Transparency and Diversity}
Participants identified strong potential for professional development, explicitly appreciating that the artifacts articulate what students \textit{``can and cannot do''} (P5). However, to maximize classroom relevance, experts recommended moving beyond black-box generation. P5 suggested adding a module to \textit{``explain the logic of level determination,''} arguing that transparency in \emph{why} a drawing was classified as "Developing" would help teachers build their own diagnostic criteria. This feedback points toward a design shift from purely generative tools to "explainable" diagnostic trainers.

\subsection{Limitations and Future Work}

Our evaluation focused on perceived plausibility and utility rather than 
learning outcomes; we do not claim that DrawSim-PD improves teacher diagnostic 
accuracy or produces measurable gains. Investigating how engagement with 
synthetic exemplars affects diagnostic reasoning development over time is an 
important next step. The framework also relies on current text-to-image 
capabilities, which show variable performance on complex topics with 
interconnected components; future iterations could explore topic-specific 
generation strategies. Finally, our evaluation prioritized corpus breadth, 
with participants collectively assessing 480 artifacts across diverse topics, 
grade bands, and performance levels. Future work includes comparative analyses 
with authentic student drawings, focused reliability studies on specific 
grade bands, and expanded validation with larger and more diverse teacher 
samples to strengthen generalizability.

%% file: sec_arXiv/conclusion_0201.tex
\section{Conclusion}

We presented \textbf{DrawSim-PD}, a generative framework that simulates 
student-like science drawings to support NGSS-aligned teacher diagnostic 
reasoning. By inverting standard generation objectives to prioritize 
pedagogical imperfection over aesthetic accuracy, we successfully modeled the 
partial understandings and spatial errors characteristic of developing learners. 
Central to our approach are \emph{capability profiles} that encode 
performance constraints, ensuring cross-modal coherence between drawings, 
narratives, and diagnostic concept maps.
In an expert-based feasibility study, K--12 science educators verified that the 
generated artifacts are well-aligned with NGSS expectations and appropriately 
differentiated by performance level, while identifying opportunities for 
refining grade-band extremes. The accompanying corpus of 10,000 artifacts 
provides open research infrastructure for calibration activities and visual 
assessment research previously constrained by privacy regulations. Beyond 
validation, we see potential for adaptive calibration systems 
targeting individual teachers' diagnostic blind spots and interactive 
generation for on-demand misconception specification.

%% file: sec_arXiv/appendix_0201.tex
\newpage
\section*{Appendix}

This supplementary material provides additional resources and examples that complement the main paper. These materials illustrate our methods concretely, explore the generated corpus in detail, and highlight differences across performance levels. Specifically, we provide:

\vspace{-2mm}
\begin{itemize}
    \item A complete \textbf{worked example} of a flowering plant life cycle diagram (aligned with NGSS 3-LS1-1) at the Advanced performance level, demonstrating the full pipeline from standards decomposition to final artifacts.
    \item \textbf{Performance level comparisons} illustrating how scientific illustrations and diagnostic concept maps evolve from Emergent to Advanced across multiple topics.
    \item \textbf{Representative examples} from the DrawSim-PD corpus demonstrating diversity across physical, life, and earth sciences.
\end{itemize}

\section{Worked Example: Life Cycle of a Flowering Plant}

This example aligns with NGSS code \textbf{3-LS1-1} (Grade 3) and illustrates \textbf{Level 4 (Advanced)} performance. It demonstrates how the system translates abstract standards into concrete generation constraints.

\subsection{Step 1: Standards Decomposition}

\begin{tcolorbox}[colback=green!5!white, colframe=green!50!black, title=Evidence Statements (Derived from NGSS 3-LS1-1), boxrule=0.4pt, arc=4pt, left=4pt,right=4pt,top=4pt,bottom=4pt]
\begin{itemize}
    \item The student can draw and label the main stages in the life cycle of a flowering plant: germination, growth, reproduction (flowering), and wilting.
    \item The student can include multiple plant types to demonstrate diversity.
    \item The student can represent the cyclic nature of life cycles, showing repetition across generations.
    \item The student can illustrate causal direction: each stage leads to the next (\emph{e.g.}, without germination, no growth).
    \item The student can explain that despite differences across plants, all life cycles share the same fundamental pattern.
    \item The student can use patterns in the life cycle to make simple predictions (\emph{e.g.}, predicting growth or flowering periods).
\end{itemize}
\end{tcolorbox}

\subsection{Step 2: Capability Profile Generation}

\begin{tcolorbox}[colback=blue!5!white, colframe=blue!50!black, title=Capability Profile: Can Do (Level 4), boxrule=0.4pt, arc=4pt, left=4pt,right=4pt,top=4pt,bottom=4pt]
\begin{itemize}
    \item The student can accurately draw and label all main stages (germination, growth, flowering, wilting).
    \item The student can include multiple plant types to show diversity.
    \item The student can clearly represent cyclic nature with arrows and repetition.
    \item The student can show how each stage causally leads to the next.
    \item The student can explain that despite visual differences, all cycles follow the same pattern.
    \item The student can use recognizable patterns to make simple predictions.
\end{itemize}
\end{tcolorbox}

\begin{tcolorbox}[colback=red!5!white, colframe=red!50!black, title=Capability Profile: Cannot Yet Do (Level 4), boxrule=0.4pt, arc=4pt, left=4pt,right=4pt,top=4pt,bottom=4pt]
\begin{itemize}
    \item The student occasionally omits less obvious labels or details when focusing on complex stages.
    \item The student inconsistently applies predictions for less familiar plant types.
\end{itemize}
\end{tcolorbox}

\subsection{Step 3: Multimodal Generation}

\begin{tcolorbox}[colback=gray!5!white, colframe=gray!50!black, title=Generated Reasoning Narrative, boxrule=0.4pt, arc=4pt, left=4pt,right=4pt,top=4pt,bottom=4pt]
I'm going to draw the life cycle of a flowering plant. First, I'll start with the seed stage where germination happens. I'll carefully draw a small seed with a sprout poking out. Next, I'll move on to growth and draw a young plant with green leaves, and I'll use arrows to show how this stage follows germination. After that, I will draw the reproduction stage with a mature plant that has bright flowers; I'll draw multiple types of plants here, like a sunflower and a rose, to show that different plants have similar life cycles. Then, I will draw the plant in the dying stage and show wilting leaves and petals. I'll make sure to label each stage clearly, and I need to remember to use arrows to show that the cycle goes on and on. I'll add labels like ``Seed,'' ``Young Plant,'' ``Flowering,'' and ``Wilting.'' While drawing, I might forget to label some smaller details, like the parts of the flower or roots, but I want to focus on getting the main stages right. Finally, I'll try to predict when the plant will flower by looking at the life cycle I've drawn, but for less common plants, I might not be too sure about the exact timing.
\end{tcolorbox}

\begin{tcolorbox}[colback=gray!5!white, colframe=gray!50!black, title=Generated Image Prompt, boxrule=0.4pt, arc=4pt, left=4pt,right=4pt,top=4pt,bottom=4pt]
A detailed and colorful life cycle diagram of a flowering plant, showing stages: seed germination, growth, flowering, and plant wilting in a cyclic pattern. Include arrows indicating the transition between stages and label each with clear text: ``Seed,'' ``Young Plant,'' ``Flowering,'' ``Wilting.'' Draw common plants such as a sunflower and a rose, showcasing diversity. Display the cyclic nature using repetitive arrows circling back to seed germination. Use clear, bright colors to make main elements distinct, with a hand-drawn school project style. Draw like a Grade~3 student, hand-drawn crayon style.
\end{tcolorbox}

\begin{tcolorbox}[colback=gray!5!white, colframe=gray!50!black, title=Prompt--Profile Alignment Check, boxrule=0.4pt, arc=4pt, left=4pt,right=4pt,top=4pt,bottom=4pt]
The image prompt captures all main elements specified in the capability profile, including life cycle stages and plant diversity. It aligns with the profile's focus on larger conceptual aspects rather than fine details (\emph{e.g.}, not requiring flower parts or roots to be labeled). The cyclic nature is emphasized, consistent with the student's ability to depict ongoing cycles and transitions, while acknowledging areas of lesser confidence such as predictions for unfamiliar species.
\end{tcolorbox}

\subsection{Step 4: Final Artifacts}

\begin{figure}[htbp]
\centering
\includegraphics[width=0.75\textwidth]{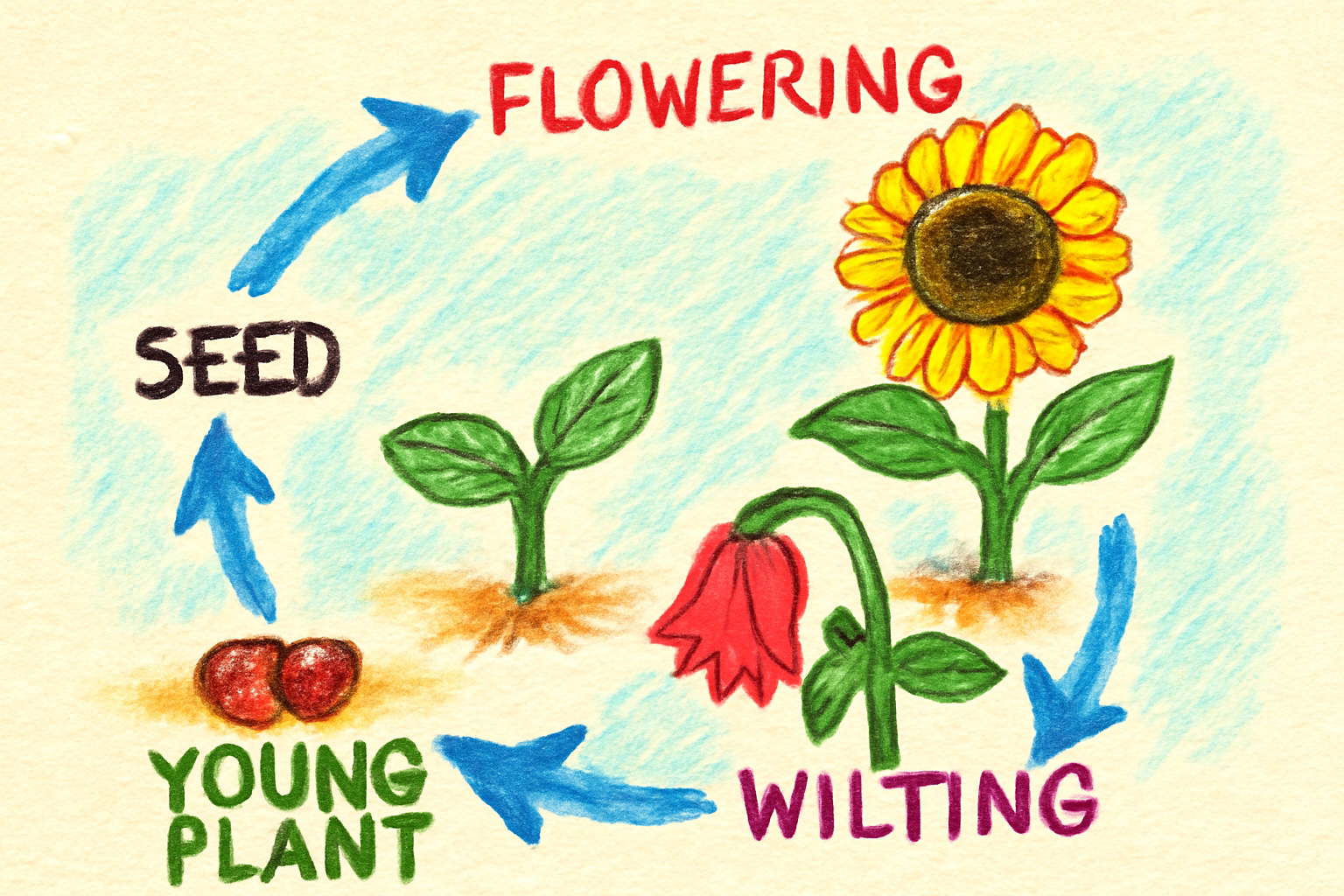}
\caption{Student-like drawing of a flowering plant life cycle (Level 4: Advanced). Note the inclusion of multiple plant types (sunflower, rose) and clear cyclic arrows, consistent with the Level 4 profile.}
\label{fig:plant_lifecycle_drawing}
\end{figure}

\begin{figure}[htbp]
\centering
\includegraphics[width=\textwidth]{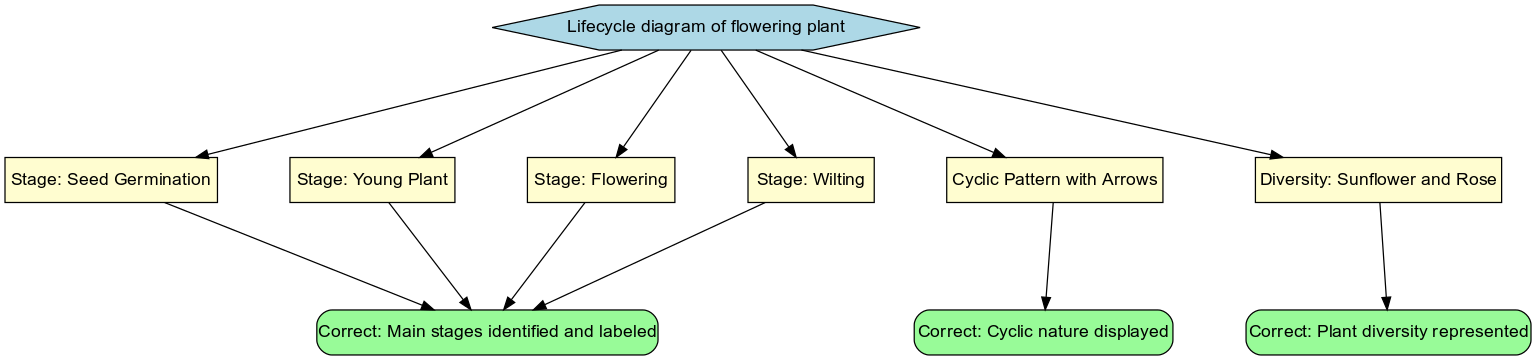}
\caption{Diagnostic concept map linking observable features in Fig.~\ref{fig:plant_lifecycle_drawing} to underlying reasoning. Note that feedback nodes (red) are absent or minimal because the student has mastered the core concepts, indicating a high proficiency level.}
\label{fig:plant_lifecycle_map}
\end{figure}

Figure~\ref{fig:plant_lifecycle_drawing} shows the rendered drawing generated from the prompt above, and Figure~\ref{fig:plant_lifecycle_map} shows the corresponding teacher-facing diagnostic concept map.

\section{Performance Level Comparisons}

This section presents comparative examples across performance levels (Emergent, Developing, Proficient, Advanced) to illustrate how student-like scientific representations vary with increasing conceptual understanding and representational competence. The comparisons highlight differences in structural completeness, causal reasoning, and use of scientific conventions. By juxtaposing illustrations aligned to the same NGSS expectations, these examples make visible the developmental progression captured by DrawSim-PD's capability profiling mechanism.

\subsection{Scientific Illustration Comparison}

\begin{figure}[htbp]
\centering
\includegraphics[width=\textwidth]{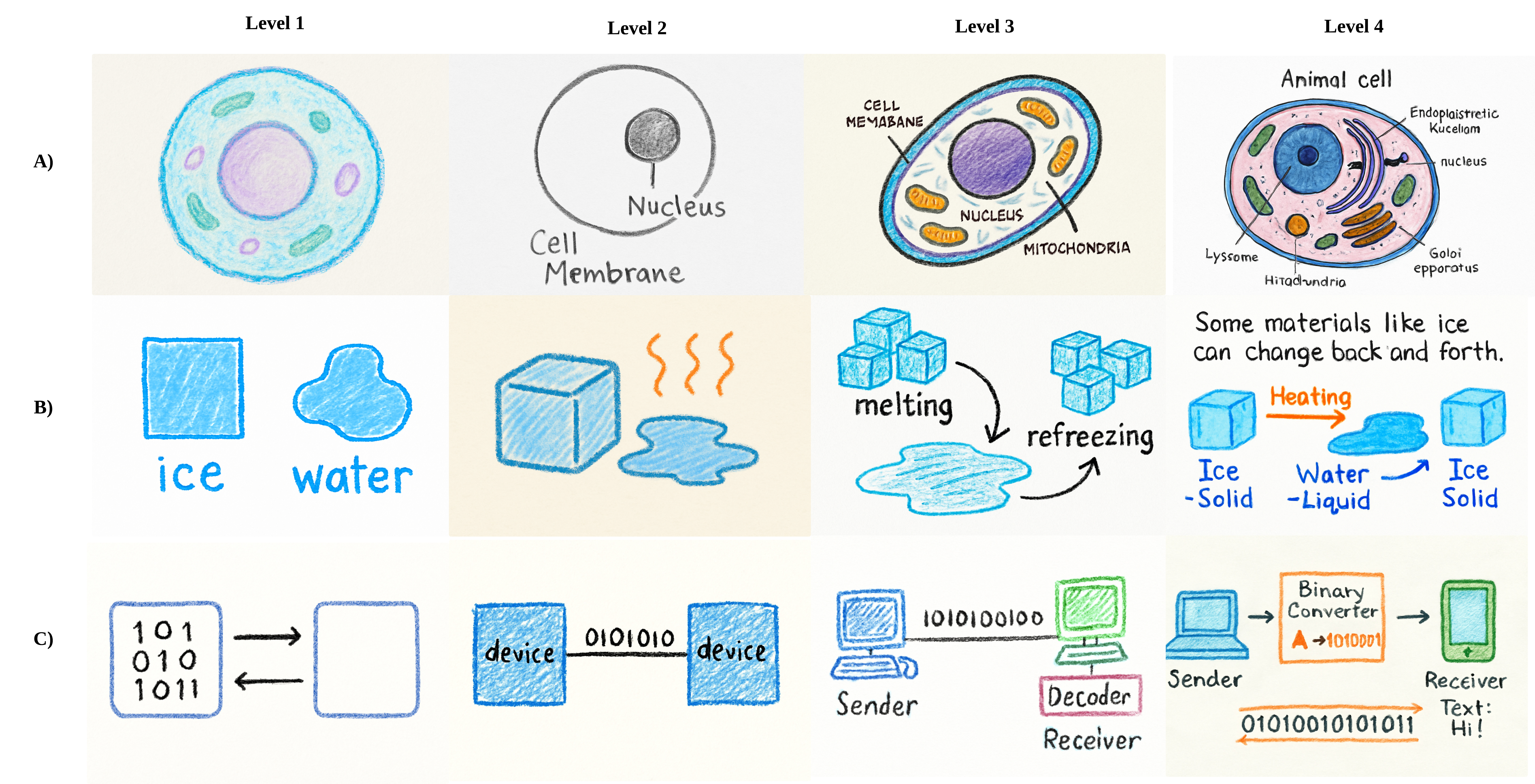}
\caption{Comparison of student-like scientific drawings across four performance levels (Emergent, Developing, Proficient, Advanced) for three NGSS topics: (A)~cell structure under a microscope (MS-LS1-1), (B)~reversible change of ice melting and refreezing (2-PS1-4), and (C)~binary code transfer system (4-PS4-3). Note the increasing detail, labeling accuracy, and structural complexity from left to right.}
\label{fig:level_comparison}
\end{figure}

Figure~\ref{fig:level_comparison} compares student-like drawings across four performance levels for three NGSS topics, illustrating developmental differences in labeling, structure, causal reasoning, and scientific conventions.

\subsection{Concept Map Comparison}

\begin{figure}[htbp]
\centering
\includegraphics[width=0.95\textwidth]{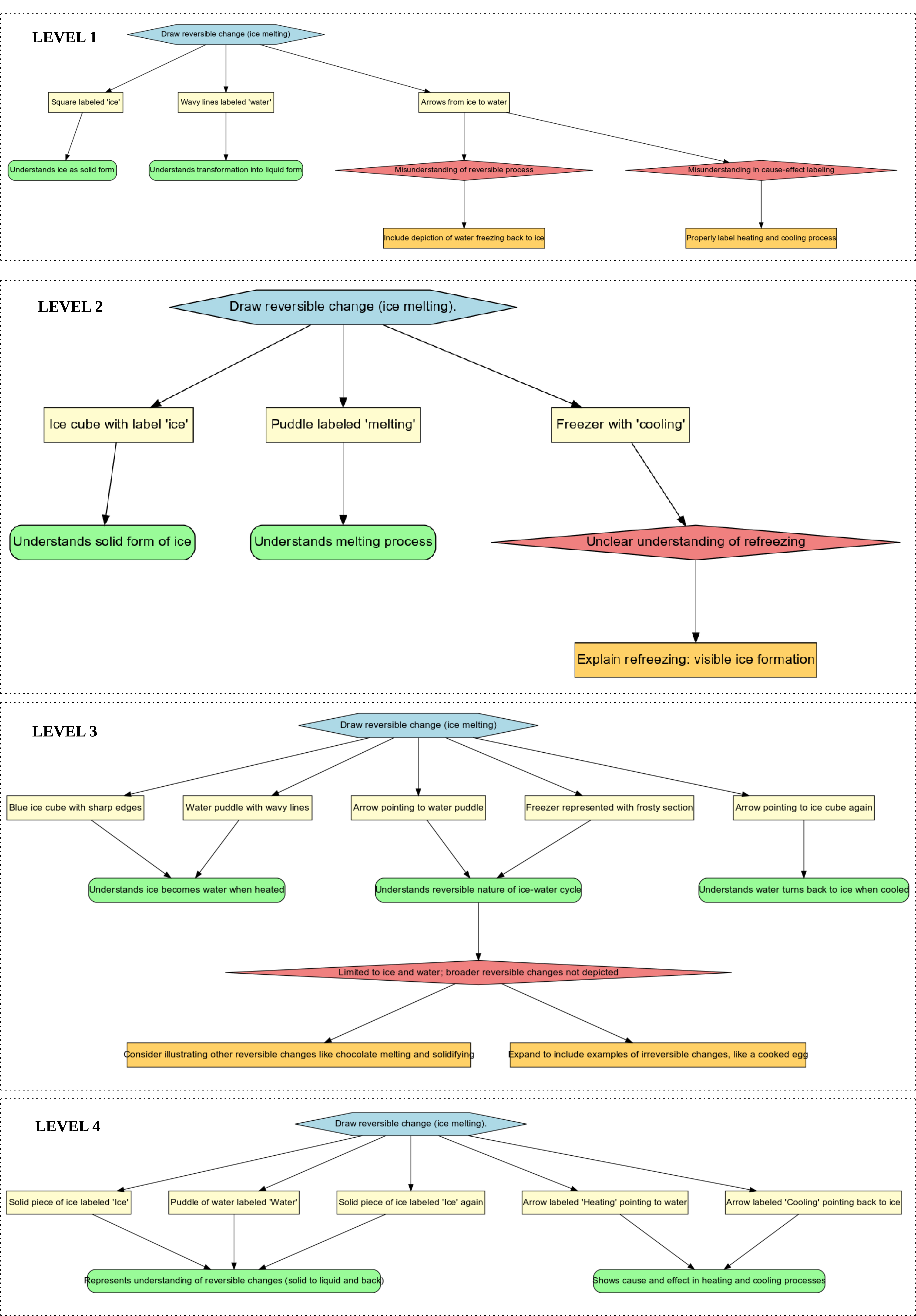}
\caption{Concept map comparison showing progression in reasoning and suggested feedback for the topic ``reversible change (ice melting)'' across four performance levels.}
\label{fig:concept_map_comparison}
\end{figure}

Figure~\ref{fig:concept_map_comparison} shows how diagnostic concept maps differ across performance levels for the same topic. Note that lower-level maps contain more ``Feedback'' nodes (red) indicating misconceptions, while higher-level maps contain more ``Understanding'' nodes (green) indicating mastery.

\newpage
\section{Representative Scientific Illustrations}

\begin{figure}[htbp]
\centering
\includegraphics[width=0.95\textwidth]{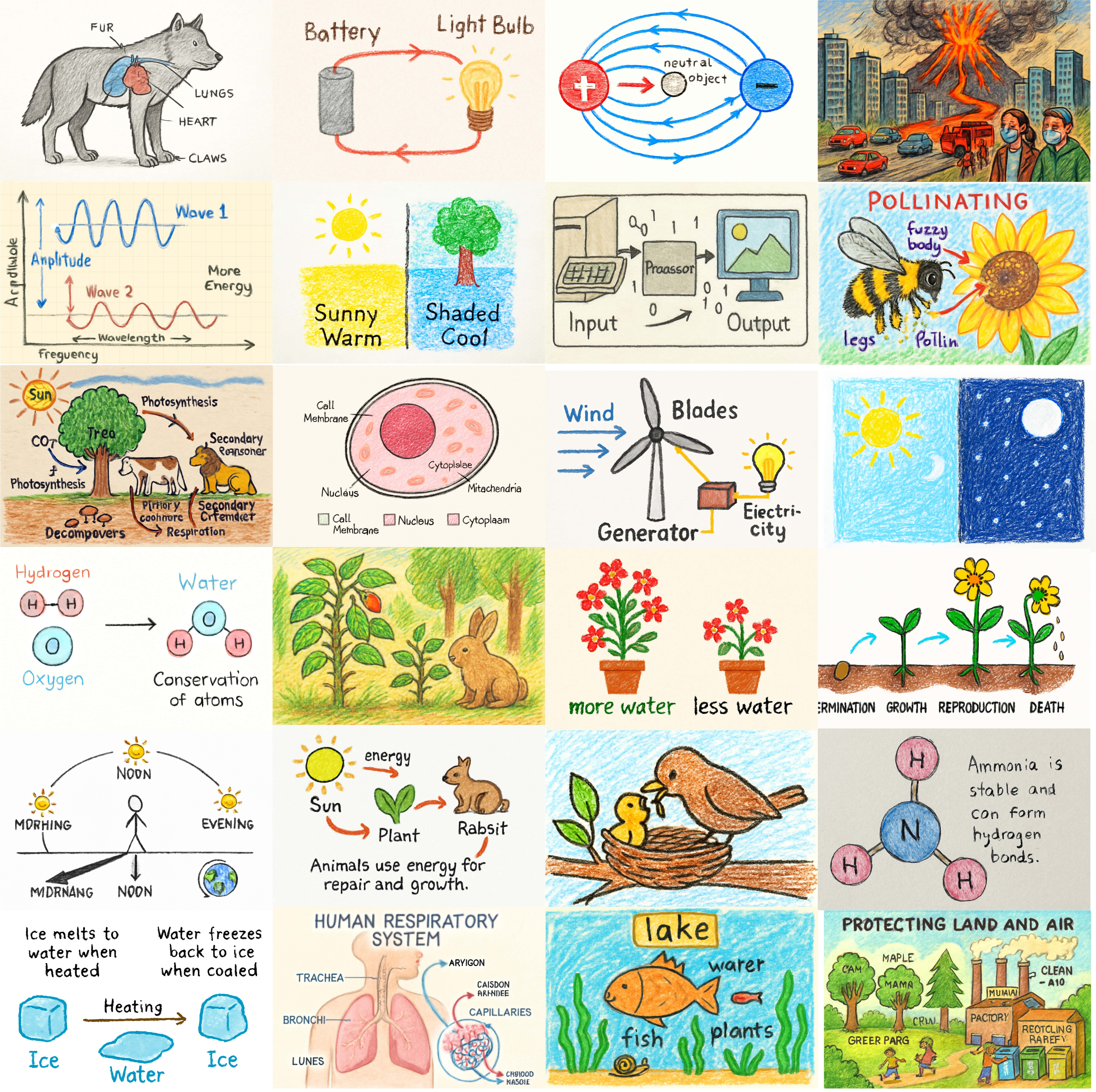}
\caption{Representative student-like scientific illustrations generated by DrawSim-PD across four performance levels (Emergent to Advanced) and three science domains (life sciences, physical sciences, earth and space sciences).}
\label{fig:dataset_examples}
\end{figure}

Figure~\ref{fig:dataset_examples} presents representative examples from the DrawSim-PD corpus, demonstrating the range of student-like artifacts across science domains and performance levels.

